\documentclass[]{IEEEtran}
\IEEEoverridecommandlockouts

\usepackage{tabularx}               
\usepackage{setspace} 
\usepackage{subfigure}
\usepackage[font={small}]{caption}
\usepackage[utf8]{inputenc}
\usepackage{booktabs}
\usepackage{epsfig} 
\usepackage{amsmath} 
\usepackage{textcomp}

\usepackage{pdfpages} 
\usepackage{dsfont} 
\usepackage{fancyhdr} 
\usepackage{setspace}
\usepackage{algorithm}
\usepackage[noend]{algpseudocode}
\usepackage{lettrine}
\usepackage{xfrac}
\usepackage{graphicx,epstopdf} 
\usepackage{mathtools}
\usepackage{epsfig} 
\usepackage{cite} 
\usepackage{amssymb}
\usepackage{multirow}
\usepackage{cases}
\usepackage{fancyvrb}
\usepackage{float}
\usepackage{atbegshi}
\usepackage{color}
\usepackage{soul}
\usepackage{steinmetz}

\usepackage{pgfplotstable}
\usepackage{array}

\usepackage{algpseudocode}
\algrenewcommand\algorithmicrequire{\textbf{Input:}}

\algrenewcommand\algorithmicprocedure{\textbf{Step:}}
\algrenewcommand\algorithmicensure{\textbf{Output:}}
\algrenewcommand\algorithmicreturn{Satisfy}
\usepackage{lipsum}

\def\BibTeX{{\rm B\kern-.05em{\sc i\kern-.025em b}\kern-.08em T\kern-.1667em\lower.7ex\hbox{E}\kern-.125emX}}

\usepackage{colortbl}
\definecolor{Gray}{gray}{0.9}
\definecolor{LightCyan}{rgb}{0.88,1,1}

\begin{document}
\title{Multi-Antenna Data-Driven Eavesdropping Attacks and Symbol-Level Precoding Countermeasures}

\author{Abderrahmane Mayouche, \IEEEmembership{Student Member, IEEE}, Wallace A. Martins, \IEEEmembership{Senior Member, IEEE}, Christos G. Tsinos, \IEEEmembership{Senior Member, IEEE},  Symeon Chatzinotas, \IEEEmembership{Senior Member, IEEE} and Bj\"orn Ottersten, \IEEEmembership{Fellow, IEEE}

\thanks{The authors are with the Interdisciplinary Centre for Security  Reliability and Trust (SnT), University of Luxembourg, L-1855, Luxembourg (E-mails: \{abderrahmane.mayouche, wallace.alvesmartins, christos.tsinos, symeon.chatzinotas, and bjorn.ottersten\}@uni.lu).

\indent This work is supported by the Engineering and Physical Sciences Research Council (EPSRC) and the Luxembourg National Research Fund (FNR) project, titled Exploiting Interference for Physical Layer Security in 5G Networks (CI-PHY), the ECLECTIC FNR project, Energy and Complexity Efficient Millimeter-Wave Large-Array Communications, and also the AGNOSTIC project with the European Research Council under Horizon 2020 grant agreement 742648. 

Part of this work has been submitted for publication in the \textit{Proceedings of the IEEE Wireless Communications and Networking Conference} (WCNC), 2021.}}

\maketitle

\begin{abstract}

In this work, we consider secure communications in wireless multi-user (MU) multiple-input single-output (MISO) systems with channel coding in the presence of a multi-antenna eavesdropper (Eve). In this setting, we exploit machine learning (ML) tools to design soft and hard decoding schemes by using precoded pilot symbols as training data. In this context, we propose ML frameworks for decoders that allow an Eve to determine the transmitted message with high accuracy. We thereby show that MU-MISO systems are vulnerable to such eavesdropping attacks even when relatively secure transmission techniques are employed, such as symbol-level precoding (SLP). To counteract this attack, we propose two novel SLP-based schemes that increase the bit-error rate at Eve by impeding the learning process. We design these two security-enhanced schemes to meet different requirements regarding complexity, security, and power consumption. Simulation results validate both the ML-based eavesdropping attacks as well as the countermeasures, and show that the gain in security is achieved without affecting the decoding performance at the intended users. 
\end{abstract}


\section{Introduction}
 While fifth generation (5G) cellular networks are currently being provisioned worldwide, its successor generation, named 6G wireless system, is being proposed to overcome several limitations in 5G \cite{6G_Saad,6G_Yang,6G_Gui}. By 2023, there will likely be $5.7$ billion total mobile users (71\% of the world population) \cite{VNI_Forecast}. In such a crowded environment, unintended receivers, e.g., an eavesdropper (Eve), may decode sensitive information given the broadcasting nature of the wireless channel \cite{PLS_6G}. As a result, security is of primary importance in next generation networks. In particular, physical-layer security (PLS) stands out as a powerful technology to complement encryption-based methods \cite{PLS_Survey}, including application-layer encryption.

The essence of PLS is to exploit the characteristics of the wireless channel, i.e., fading, noise, interference, and diversity, to attain an acceptable decoding performance at intended users while obstructing the correct decoding at Eve. Alternatively, the aim of PLS is to increase the gap of correct decoding rates between intended users and Eve \cite{PLS_jehad}. PLS is foreseen to be used as a complementary layer of protection, in addition to the existing cryptography-based security methods. As the rise of quantum computing \cite{Quantum_Survey,Crypto_Quantum} is threatening both symmetric and asymmetric cryptography, non-cryptographic-based methods such as PLS are needed \cite{SLP_PLS_IoT}. 
 
In this context, symbol-level precoding (SLP) \cite{SLP_Survey_2020,Maha_Survey,Masouros_1st} has been introduced as a new way for attaining PLS \cite{Ashkan}. Although not originally conceived as a PLS method, SLP is more secure than block-level precoding, like zero-forcing (ZF) \cite{ZF_Andrea}, as the precoder is redesigned for each symbol period. In \cite{SLP_PLS_1,SLP_PLS_2}, secure SLP precoding schemes were proposed in the context of a multiple-input single-output (MISO) wiretap channel while considering only a single-antenna Eve. In \cite{Smart_Eve}, the authors proposed to exploit the statistical characteristics of the received signal at Eve in order to improve its detection performance. To counter this vulnerability, the authors in \cite{Smart_Eve} proposed secure SLP-based precoding schemes to degrade Eve's performance. 

Besides, machine learning (ML) has attracted significant interest in the area of wireless communications \cite{ML_WN}. By definition, machine learning is a core subset of artificial intelligence (AI), which is an ensemble of tools and algorithms intended for making predictions or decisions through learning patterns from data \cite{nextgen_ML}. In other words, based on a dataset, ML algorithms build a mathematical model in order to make predictions or decisions. 

AI has been envisioned by several researchers as the most prominent feature of 6G \cite{6G_AI}, since it is an efficient tool for several contemporary complex scenarios. For instance, ML techniques can be categorized into two distinct objectives related to extracting patterns from data: first, performance improvement, in which ML is used to optimize the operating parameters at the lower layers; second, information processing of the huge data generated by wireless devices at the application layer \cite{ML_Categories}. 

Nevertheless, the potential of ML is not fully exploited in PLS, although ML for PLS has been explored in some recent works \cite{Def_Mod_det,Karl_autoencoder,DL_Wtap_design,DL_Wtap_design}. In \cite{Def_Mod_det}, the authors proposed an attack where ML is used to determine the underlying modulation scheme. In \cite{Karl_autoencoder}, the authors employed ML for wiretap code design considering Gaussian channels under finite block length, and a similar idea was proposed in \cite{DL_Wtap_design}. 

In the context of multi-user (MU) MISO systems, a related PLS work is \cite{Abdu_OJ_COMs}, where we proposed an ML-based attack in uncoded systems, in which an Eve can use ML to improve its detection performance via pilot symbols. Since most communication systems employ forward-error correction (FEC), it is important to investigate eavesdropping in systems that feature FEC. To that end, in our present work, we go beyond \cite{Abdu_OJ_COMs} by considering a practical scenario where channel coding is employed. It is worth mentioning that the Eve can exploit the redundancy induced by channel coding to improve its decoding capabilities during the attack. 

Herein, in a FEC-enabled MU-MISO system with a multi-antenna Eve, we first propose ML frameworks that allow an Eve to soft/hard decode the transmitted message with good accuracy, i.e., coded FER at Eve around $10^{-3}$. After introducing these two decoding attacks, we validate them in the aforementioned MU-MISO system, and show that even conventional SLP-based schemes \cite{Ashkan} are vulnerable to such attacks. As a countermeasure to these attacks, we propose two novel security-enhanced SLP-based schemes that impair the ML training process, thus enhancing security. Simulation results show the efficacy of the ML-based attacks against conventional precoders, i.e., very low bit-error rate (BER) at Eve, indicating good decoding performance, and the effectiveness of the proposed countermeasures, i.e., high BER at Eve even with numerous antennas, implying poor decoding performance. The primary contributions of the paper are listed below:

\begin{enumerate}
    \item We introduce eavesdropping attacks in MU-MISO systems with FEC by proposing novel ML-based soft and hard decoding schemes, where a multi-antenna Eve can use ML and the knowledge of pilot symbols as well as the added redundancy related to channel coding to decode the transmitted data with high accuracy.   
	\item We introduce the soft decoding scheme by proposing an ML framework that can be used by an Eve to correctly soft-decode messages sent to a particular user in an MU-MISO system. Furthermore, we also propose an ML-based hard decoding scheme at the Eve. We design the ML framework of this scheme to directly predict the coded bits. 
	\item To counteract these eavesdropping attacks, we propose two security-enhanced SLP-based schemes that aim to increase the BER at Eve by impeding the learning process. This is performed by either embedding randomness in Eve's received signal or minimizing Eve's received power. We note that the proposed schemes assume perfect knowledge of Eve's channel at the BS \cite{Masouros}, which is the case when Eve is part of the system trying to eavesdrop other users.
	\item We design these two PLS schemes in such a way that different requirements for security, complexity, and power consumption are met, to provide the base station (BS) with options to choose the most suitable scheme depending on the desired criteria. 
	\item We validate the eavesdropping attacks as well as the countermeasures through extensive simulations, where we show the vulnerability when using non-secure precoding schemes and the drastic security gains when using our proposed PLS schemes.

\end{enumerate}

The rest of the paper is organized as follows: Section \ref{SM} describes the system model. In Section \ref{ML-A}, we introduce the ML-based attacks, whereas in Section \ref{CM}, we propose our novel SLP-based schemes as countermeasures to this attack. Simulation results are discussed in Section \ref{NR}, followed by the conclusion in Section \ref{CC}.

\textbf{\textit{Notations:}} $\lVert \cdot \lVert$ represents the Euclidean norm. $\mathbb{R}^{m\times n}$ and $\mathbb{C}^{m\times n}$ represent the set of ${m\times n}$ real matrices, and the set of ${m\times n}$ complex matrices, respectively. The superscript $(\cdot)^{\intercal}$ the transpose operator, whereas ${\rm Re}\{\cdot\}$ and ${{\rm Im}\{\cdot\}}$ denote the real and the imaginary parts of a complex number. Upper and lower boldface symbols are used to denote matrices and column vectors, respectively.
  
\section{System model} \label{SM}

\begin{figure}[t]
\centering
  \includegraphics[width=3in]{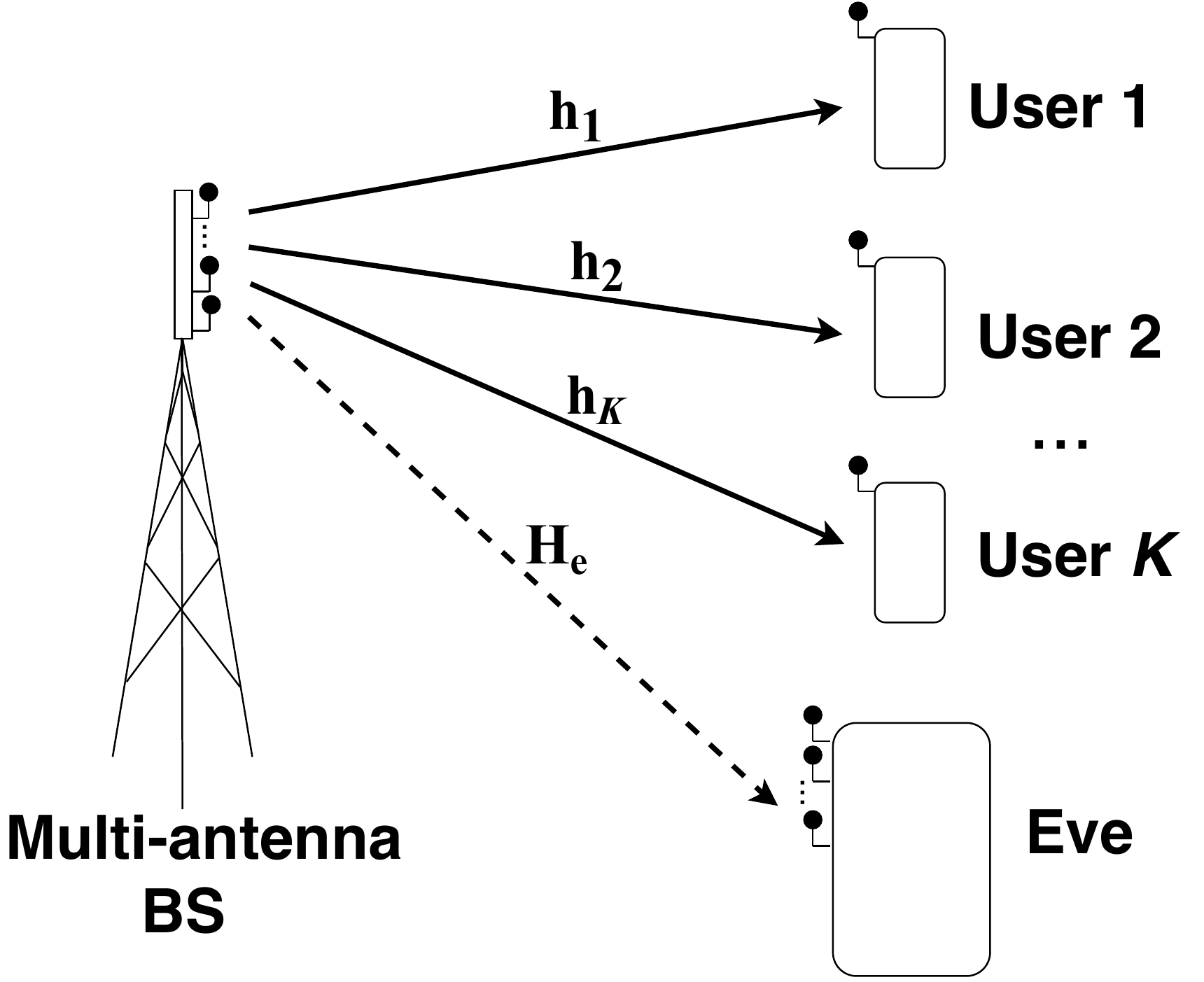}\\ 
  \caption{Downlink MU-MISO system comprised of: a BS with $N_{\rm t}$ antennas, $K$ single-antenna users, and one Eve with $M$ antennas.}
  \label{Dl_MU-MISO_system}
\end{figure} 

As depicted in Fig.~ \ref{Dl_MU-MISO_system}, we consider a single-cell MU-MISO downlink system, where the BS is equipped with $N_{\rm t}$ transmit antennas serving $K$ single-antenna users, with $K \leq N_{\rm t}$, and one multi-antenna Eve with $M$ antennas. We assume a block fading channel $\mathbf{h}_k \in\mathbb{C}^{1 \times N_{\rm t}}$ between the transmit BS antennas and the $k$-th user. The received coded signal by the $k$-th user at the symbol slot $n$ can be expressed as

\begin{equation} 
{y_k[n] =  \mathbf{h}_k \mathbf{x}_{\rm d}[n] + z_k[n]},
\end{equation}

\noindent where $\mathbf{x}_{\rm d}[n] \in\mathbb{C}^{N_{\rm t}\times 1}$ is the transmitted coded vector from the $N_{\rm t}$ transmit antennas, and $z_k[n] \in\mathbb{C}$ is the additive white Gaussian noise (AWGN) at the $k$-th user with variance $\sigma^2_z$. 

The above model can be rewritten in a matrix form by collecting the received signal at all users in vector $\mathbf{y}[n] \in\mathbb{C}^{K \times 1}$ as

\begin{equation} 
{\mathbf{y}[n] =  \mathit{\mathbf{H}} \mathbf{x}_{\rm d}[n] + \mathbf{z}[n]}, 
\end{equation}

\noindent where $\mathit{\mathbf{H}} = [\mathbf{h}_1^{\intercal} \cdots \mathbf{h}_K^{\intercal}]^{\intercal} \in\mathbb{C}^{K \times N_{\rm t}}$ represents the system channel matrix and $\mathbf{z}[n] \in\mathbb{C}^{K \times 1}$ collects the independent AWGN components of all users. 

Similarly, the received signal at Eve, $\mathbf{y}_{\rm e}[n] \in\mathbb{C}^{M \times 1}$, can be expressed as follows:

\begin{equation} 
{\mathbf{y}_{\rm e}[n] =  \mathit{\mathbf{H}}_{\rm e} \mathbf{x}_{\rm d}[n] + \mathbf{z}_{\rm e}[n]}, 
\end{equation}

\noindent where $\mathit{\mathbf{H}_{\rm e}} = [\mathbf{h}_{{\rm e},1}^{\intercal} \cdots \mathbf{h}_{{\rm e},M}^{\intercal}]^{\intercal} \in\mathbb{C}^{M \times N_{\rm t}}$ represents the system channel matrix between the BS and the multi-antenna Eve, and $\mathbf{z}_{\rm e}[n] \in\mathbb{C}^{M \times 1}$ assembles the independent AWGN components at the $M$ antennas, with a variance of $\sigma^2_{\rm e}$ each. 

We note that the pilot symbols, also being referred to as reference signals, are an integral part of communication systems that are known entities to all parties. In particular, they are commonly used for channel-state information (CSI) and signal-to-interference-plus-noise ratio (SINR) estimation. Specifically, non-precoded pilot symbols are used for CSI estimation while precoded pilot signals are intended for SINR estimation \cite{ETSI_DVB}. In this work, we are interested in the latter case, precoded pilot symbols, which uses the same modulation and coding scheme (MCS) used for precoding the data. In this context, we define $N$ as the number of precoded pilot symbols used within a frame. We also note that these $N$ pilot symbols are interleaved with data symbols in a frame that fits within the channel coherence time $T$. In this setting, we define the input data symbols intended for the $K$ users as $\mathbf{d} \in\mathbb{R}^{K \times 1}$, with $d_k$ being the symbol intended for user $k$.

In the case of block-level precoding, we define $\eta$ as the mean power. For the SLP case, we define $\gamma_k$ as the target SINR for the $k$-th user with $\boldsymbol{\gamma} = [\gamma_1 \ldots \gamma_K] \in\mathbb{R}^{K \times 1}$ representing the target SINR for all users. For ease of notation, we drop the time index $n$ in the remainder of the paper.

\section{ML-based attacks}\label{ML-A}

In this section, we will propose two ML eavesdropping attacks, where a multi-antenna Eve uses precoded pilot symbols as training data to  accurately hard/soft decode the transmitted symbols. We start by presenting the motivation of our work. Next, we present the ML frameworks for the proposed soft and hard decoding schemes. 

\subsection{Motivation}\label{Motiv}

To motivate our work, we study the received signal at Eve when the BS sends precoded pilot signals to the intended users. First, we investigate the case when the BS uses a conventional block-level precoder, i.e.,  ZF \cite{ZF_Andrea}. Next, we examine the case of a conventional SLP precoder, i.e., the constructive interference for sum power minimization (CISPM) approach in \cite{Maha_TSP_15}. 
 
For illustration purposes, we consider the following example: an MU-MISO system with $N_{\rm t} = 15$, $K = 6$, $\sigma_z^2 = 1$, one channel realization, and quadrature phase-shift keying (QPSK) as a modulation scheme, where the BS sends to each user two precoded pilot signals of $N = 150$ symbols each. In this setting, a single-antenna Eve attempts to eavesdrop a specific user $k$.

\begin{figure}[t]
\centering
  \includegraphics[width=3.5in]{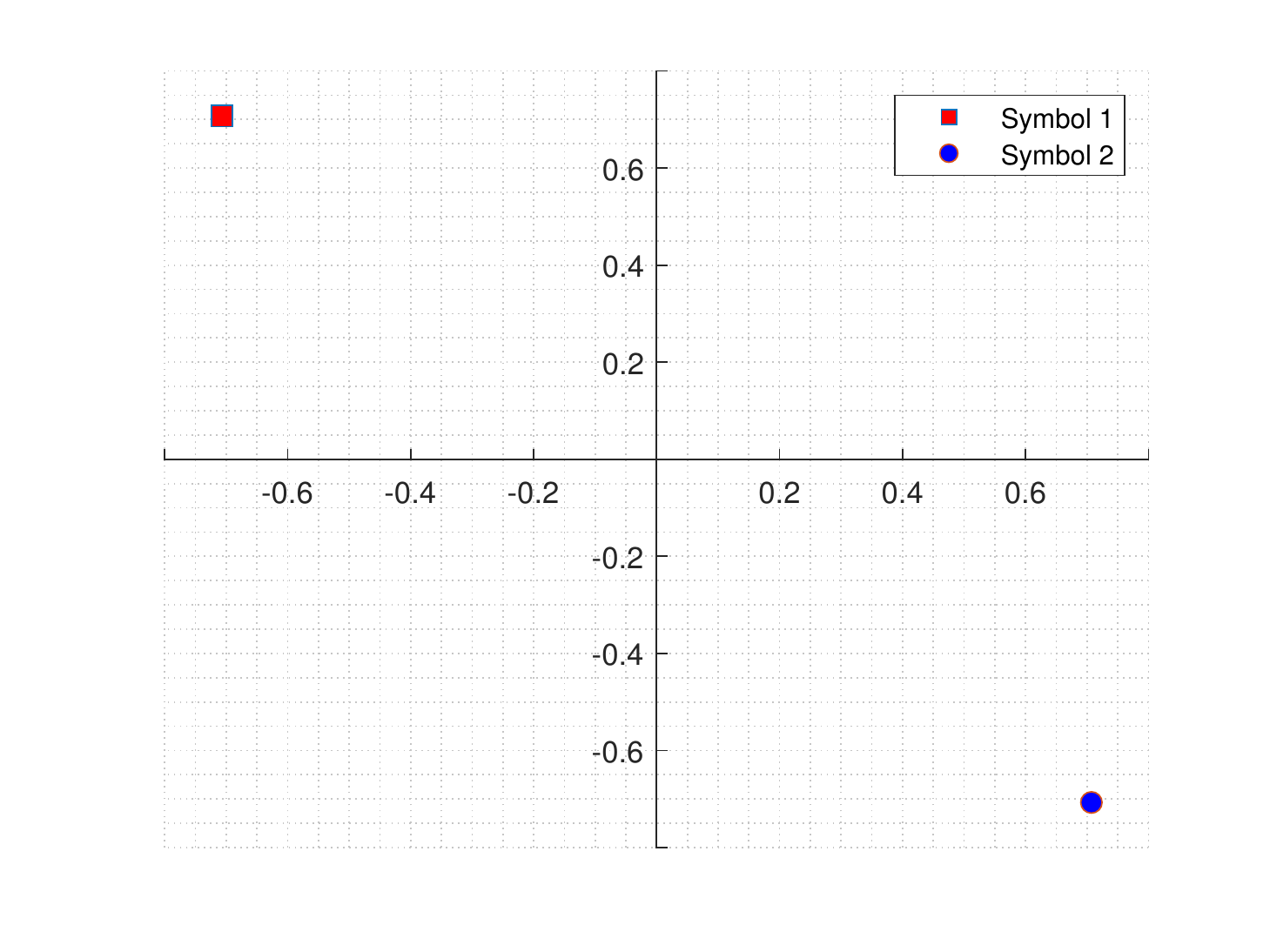}\\ 
  \caption{Pilot symbols constructing the two pilot signals intended for user $k$.}
  \label{tx_signals_u_k}
\end{figure}

To better understand the example visually, the BS sends the same symbols to user $k$ while it sends pseudo-random sequences to the remaining users. The two symbols constructing the two pilot signals intended for user $k$ are plotted in Fig.~\ref{tx_signals_u_k}.

\begin{figure}[htbp]
\centering

\begin{subfigure}
\centering
\includegraphics[width=\linewidth]{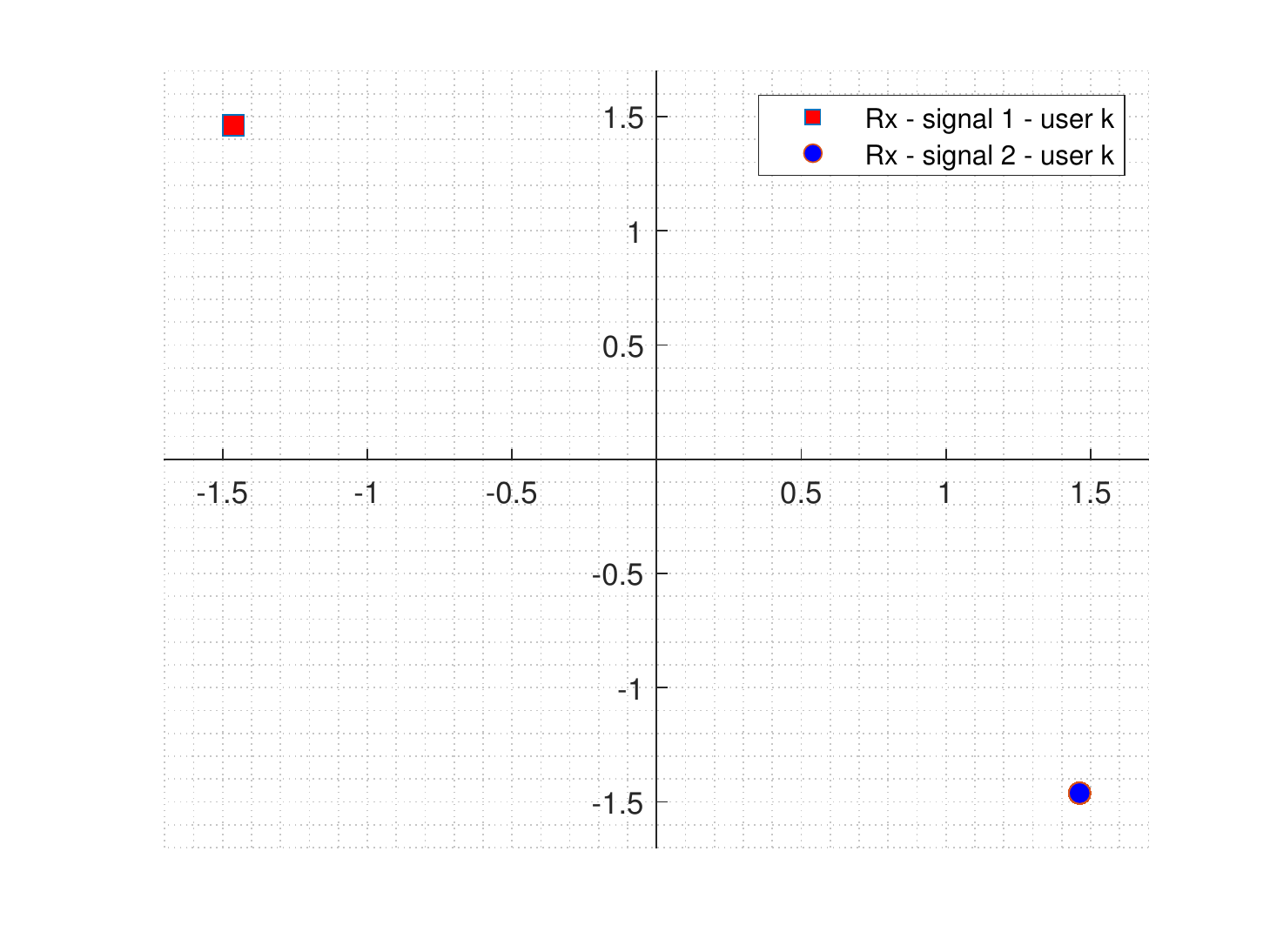}
\caption*{(a) Noiseless received signals at user $k$}
\label{fig:subim2}
\end{subfigure}

\begin{subfigure}
\centering
\includegraphics[width=\linewidth]{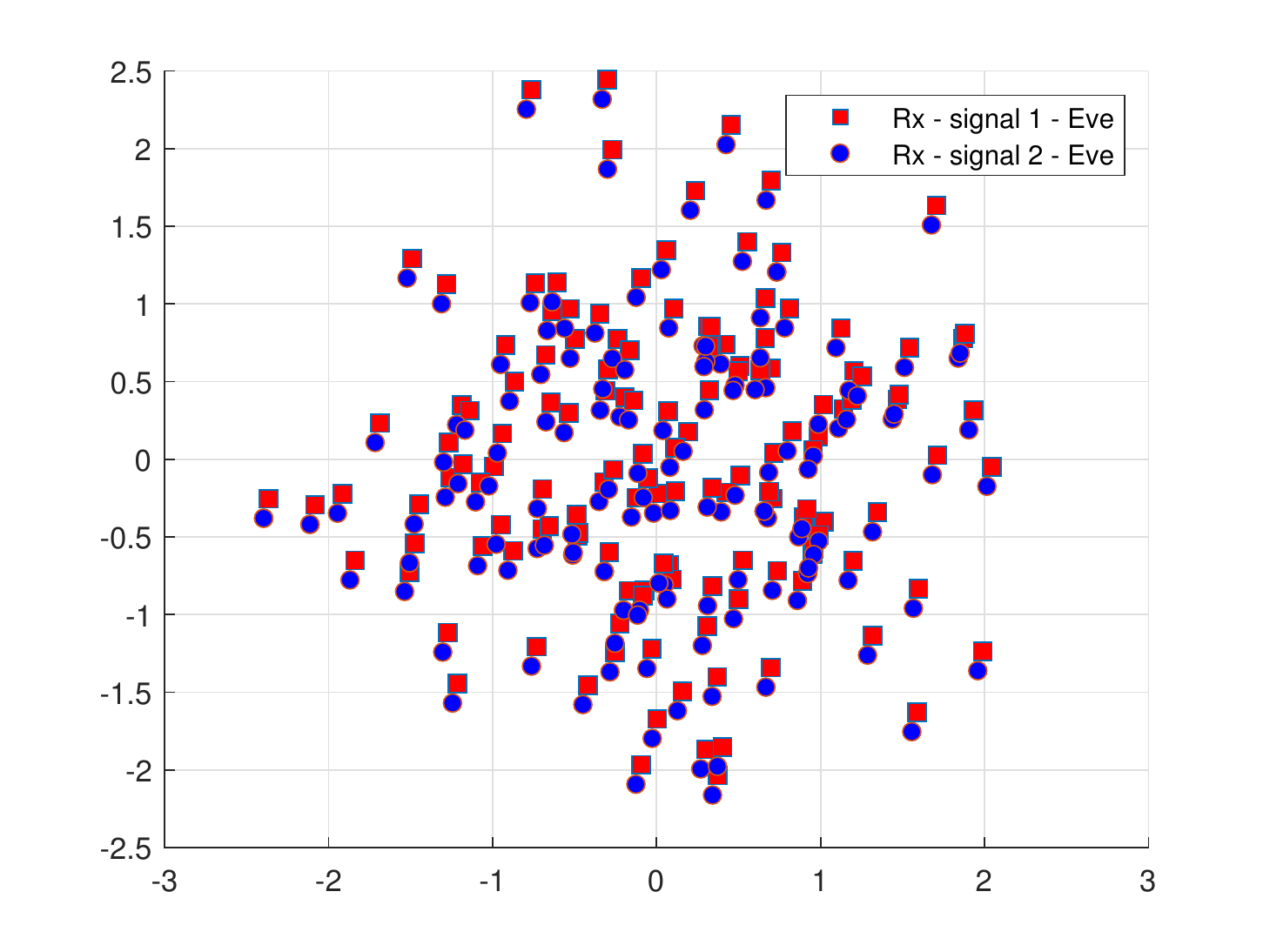}
\caption*{(b) Noiseless received signals at Eve}
\label{fig:subim2}
\end{subfigure}

\caption{Noiseless received signals at user $k$ and Eve when the BS uses ZF precoding with a mean power of $ 5~\mathrm{dB}$. }
\label{fig:image2}
\end{figure}

When the BS precodes the aforementioned pilot signals with ZF precoding of mean power of $ 5~\mathrm{dB}$, the noiseless received signals at user $k$ and the Eve are respectively given as in Fig.~\ref{fig:image2}(a) and Fig.~\ref{fig:image2}(b), respectively. We note that the channel to the $K$ users and Eve were generated randomly.

As depicted in Fig.~\ref{fig:image2}(a), the received signal at user $k$ shows no inter-user interference as it was cancelled by the ZF precoder. However, the received signal at Eve is spread due to the inter-user interference effect, as Eve's channel is different from user's $k$ channel. Still, we can observe a precoding pattern that applies to both received signals, i.e., the red squares are mostly positioned on the top right of the blue circles.

A more inherently-secure precoding scheme, which does not depict patterns of the precoding used, is SLP precoding \cite{Maha_TSP_15}. This particular SLP scheme is designed to exploit the multi-user interference for power gains at the intended users. In other words, this scheme propels the intended users' received signals deeper into the correct detection region of the desired symbol for each intended user. The corresponding optimization problem is defined as 

\noindent
\begin{IEEEeqnarray}{rCl} \label{SLP_PLS}
\mathbf{x}_{\rm d} (\mathbf{d}, \mathbf{H},\boldsymbol{\gamma}) &{=}& 
\displaystyle \arg \min_{\mathbf{x}}  {||\mathbf{x}||^2 }  \\
&& \text{subject to} \nonumber\\ 
&& {\rm Re}\{\mathbf{h}_k \mathbf{x} \} \trianglelefteq \sigma_z \sqrt{\gamma_k} {\rm Re}\{d_k\}, \; \forall k \nonumber\\
&& {\rm Im}\{\mathbf{h}_k \mathbf{x} \} \trianglelefteq \sigma_z \sqrt{\gamma_k} {\rm Im}\{d_k\}, \; \forall k \nonumber,
\end{IEEEeqnarray}
\noindent where the operator $\trianglelefteq$ denotes the correct detection region \cite{SLP_16}.

 \begin{figure}[t]
\centering
  \includegraphics[width=2in]{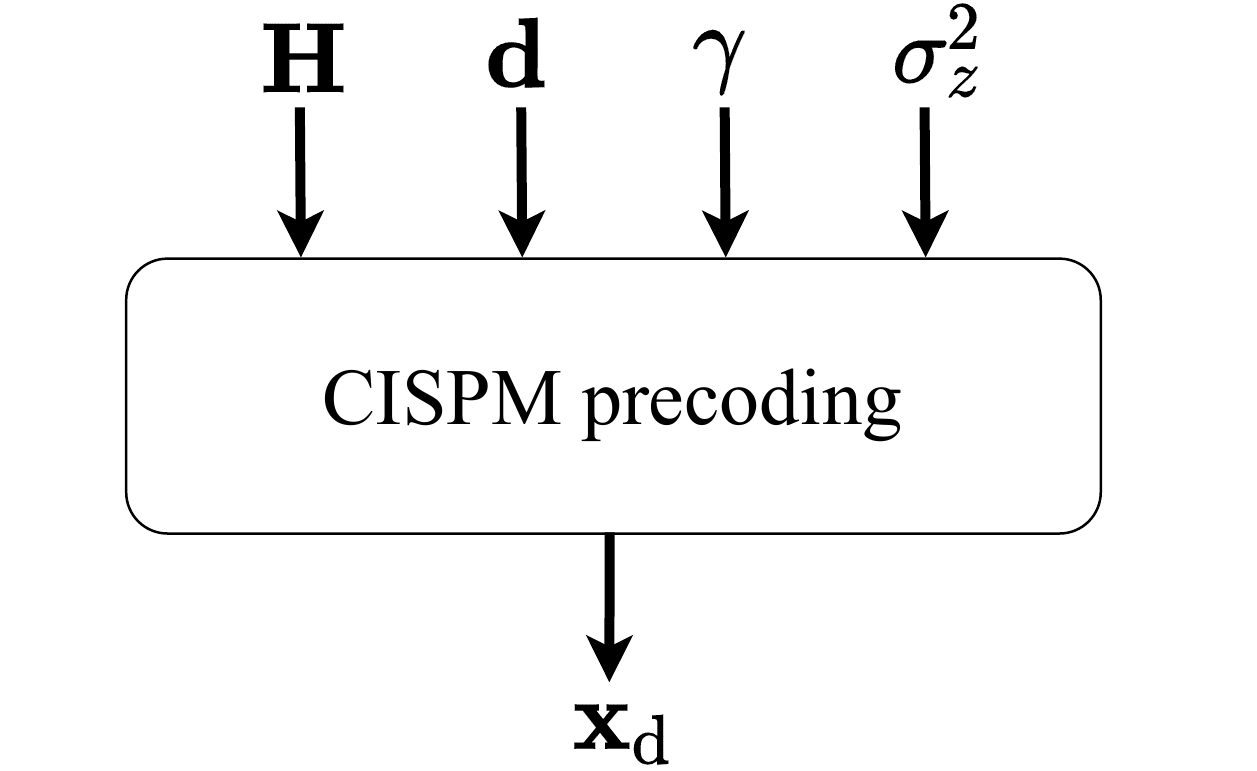}\\ 
  \caption{CISPM precoding scheme.}
  \label{CISPM_d}
\end{figure} 

As depicted in Fig.~\ref{CISPM_d}, the CISPM precoding takes inputs: the channel to the intended users, $\mathbf{H}$, the input data to be transmitted to the intended users, $\mathbf{d}$, the target SINR for all intended users, $\boldsymbol{\gamma}$, and the noise variance at the users $\sigma^2_z$. 

\begin{figure}[htbp]
\centering

\begin{subfigure}
\centering
\includegraphics[width=\linewidth]{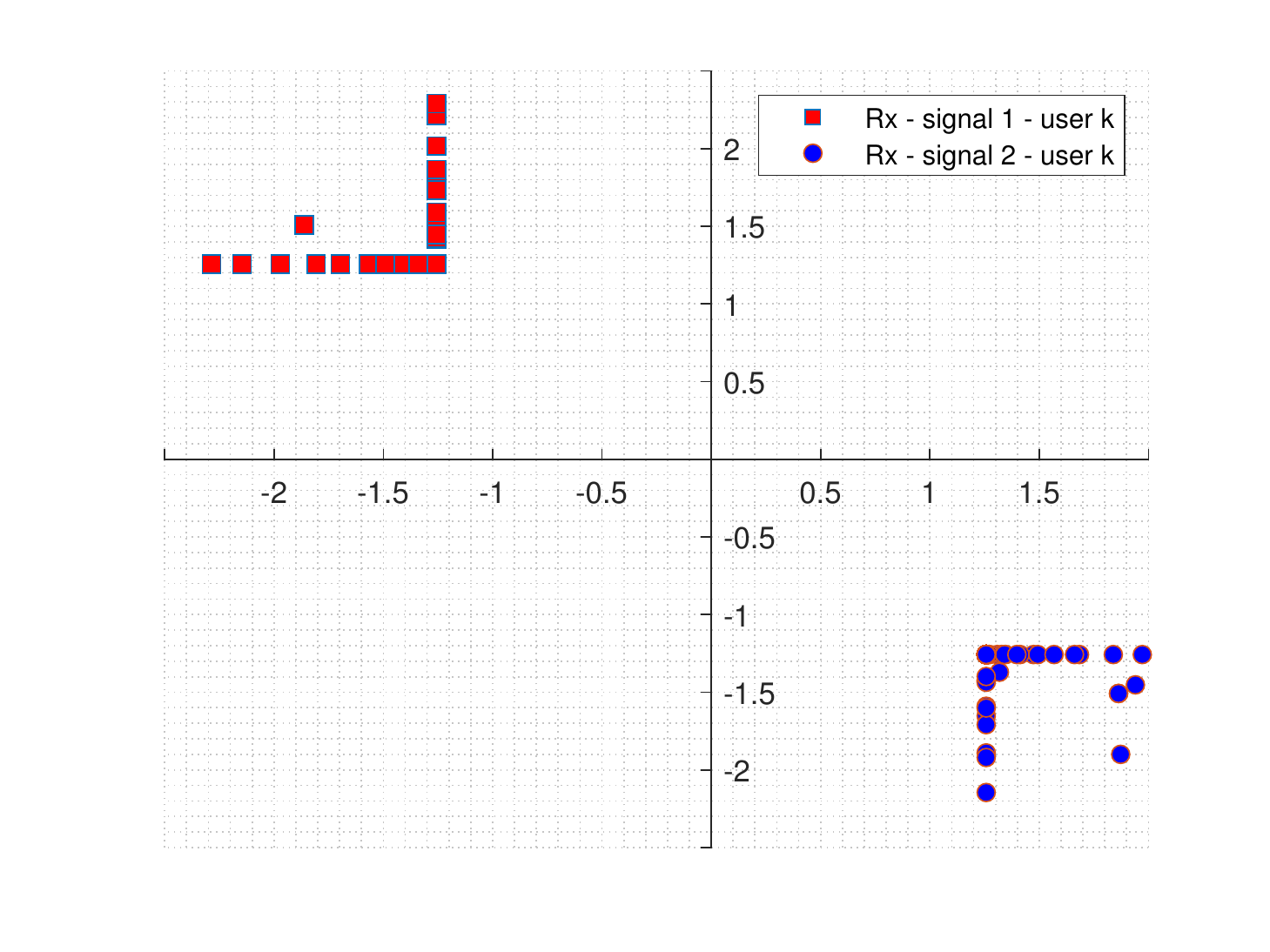}
\caption*{(a) Noiseless received signals at user $k$}
\label{fig:subim2}
\end{subfigure}

\begin{subfigure}
\centering
\includegraphics[width=\linewidth]{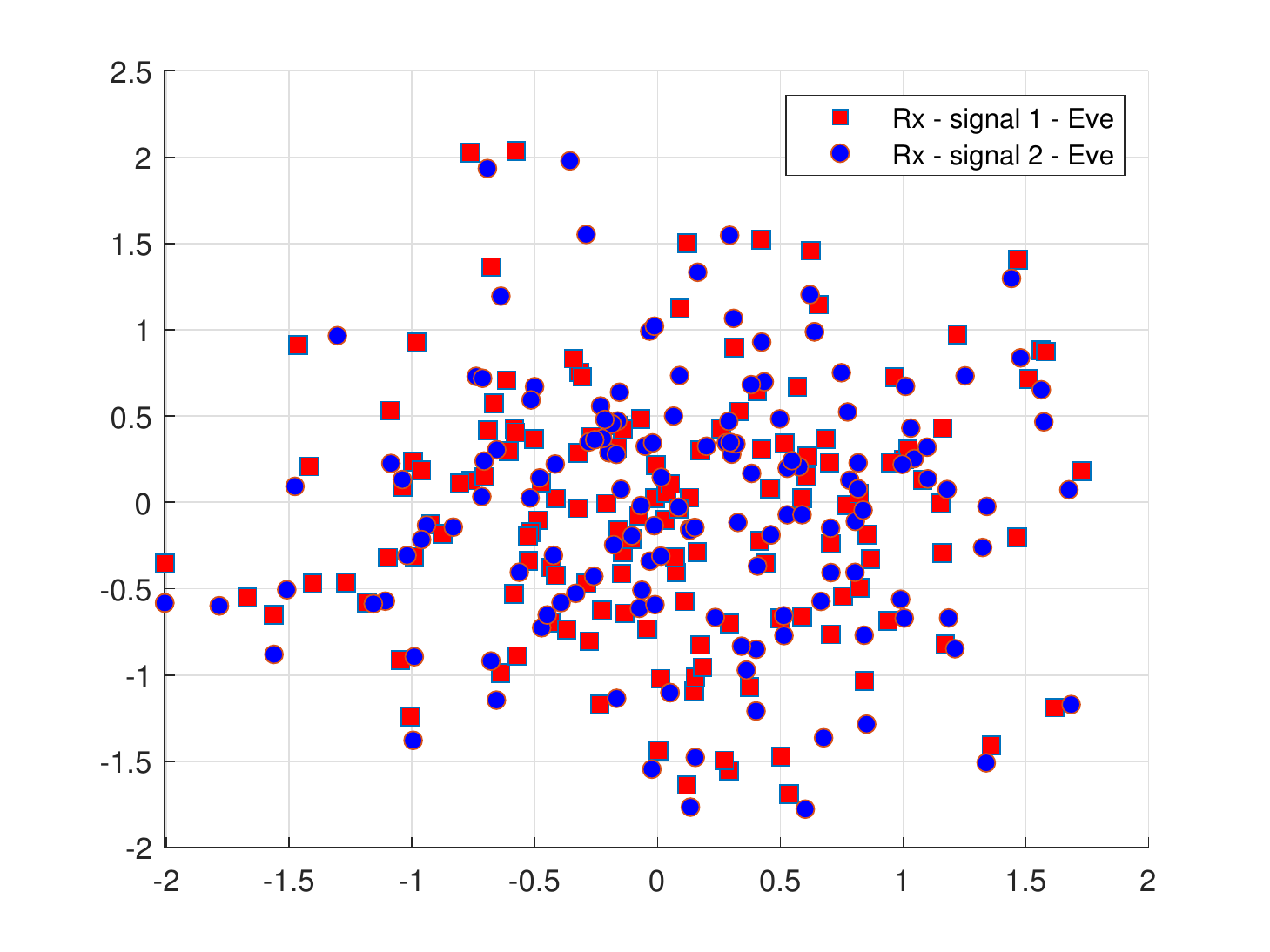}
\caption*{(b) Noiseless received signals at Eve}
\label{fig:subim2}
\end{subfigure}

\caption{Noiseless received signals at user $k$ and Eve when the BS uses CISPM precoding. }
\label{fig:image3}
\end{figure}

Now we consider the aforementioned toy example of transmitted pilot signals illustrated in Fig.~\ref{tx_signals_u_k}, but with the BS using CISPM precoding with a target SINR value of $ 5~\mathrm{dB}$ for each user as represented in Fig.~\ref{CISPM_d}. The corresponding results of this example are illustrated in Fig.~\ref{fig:image3}. 

 Fig.~\ref{fig:image3}(a), represents the noiseless received signal at user $k$. As expected when using SLP precoding, the inter-user interference is transformed into power gains, resulting in deviations of the received signal deeper into the detection region while guaranteeing a specific target SINR value. The noiseless received signal at Eve when the aforementioned signals are transmitted is plotted in Fig.~\ref{fig:image3}(b). As opposed to Fig.~\ref{fig:image2}(b) where the blue circles are always below and to the left of the red boxes, in Fig.~\ref{fig:image3}(b) there is no fixed pattern.

This discrepancy is due to the randomness of the input data to transmit $\mathbf{d}$, which changes at each symbol slot resulting in $\mathbf{x}_{\rm d}$ to vary accordingly. Even though Eve is trying to eavesdrop user $k$ whose received symbols do not change, what Eve receives provides no apparent insights about what was transmitted.  

\begin{figure}[t]
\centering
  \includegraphics[width=3.5in]{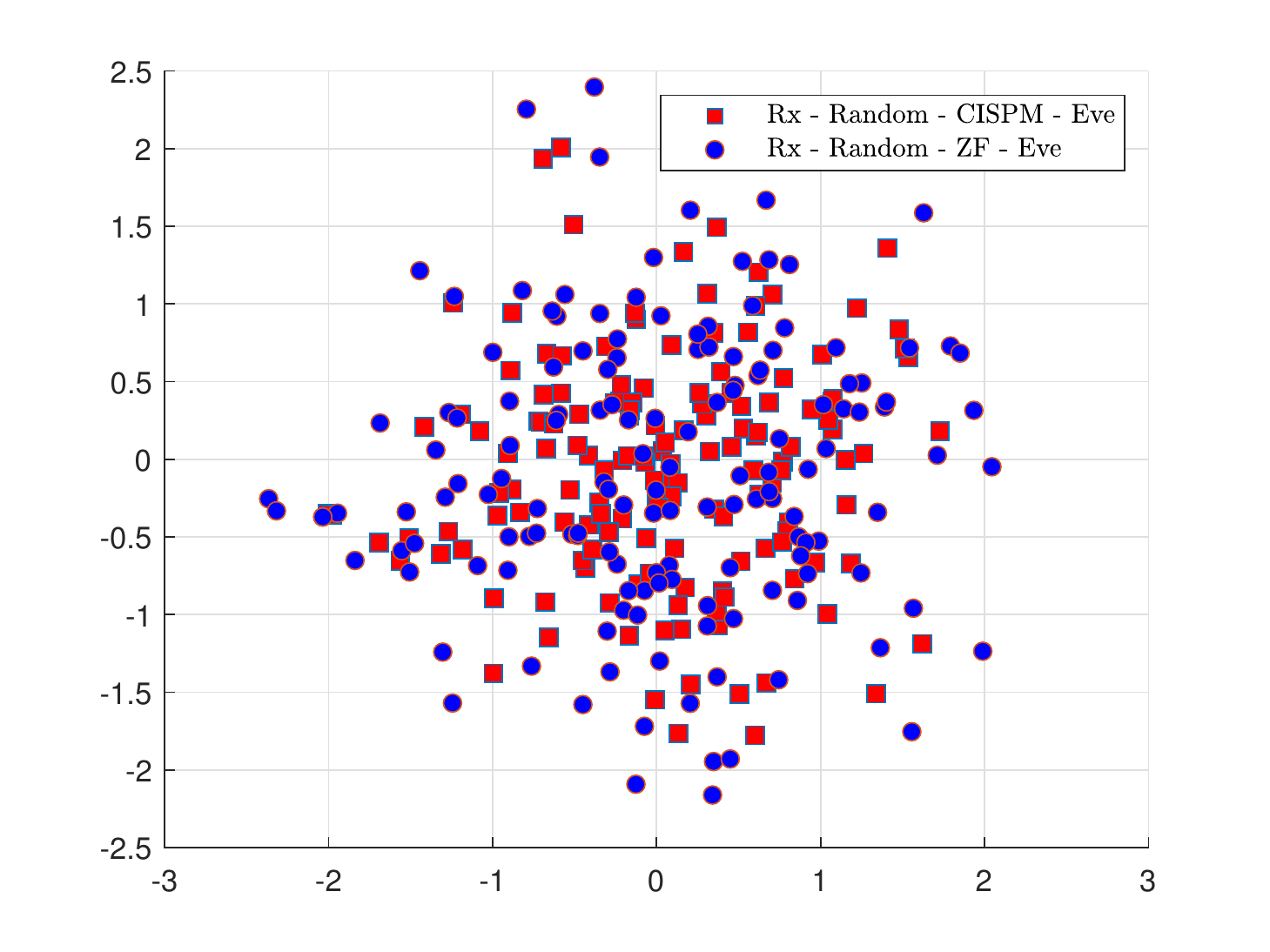}\\ 
  \caption{Received signal at Eve when BS sends pseu-random sequences to every user.}
  \label{rx_random_C_Z}
\end{figure} 

Naturally, when the BS sends the actual precoded pilots, as demonstrated in Fig.~\ref{rx_random_C_Z}, then the received signal will exhibit more randomness when compared to the example in Fig.~\ref{tx_signals_u_k}, as the actual pilots are pseudo-random sequences for all users. Hence, using conventional detection techniques to directly decode the received signal at Eve will result in a poor performance. 

Thus, in this work we propose to use ML to model the non-linear mappings underlying this apparent randomness, so that Eve can decode with an acceptable BER. Since the symbols used for the precoded pilots are known in communication standards to all parties, we propose to use ML to leverage this knowledge and decode the transmitted data to a particular user with decent accuracy. To that end, we propose ML-based soft and hard decoding schemes that can accurately decode the transmitted signal by using the precoded pilot symbols as training data. For a thorough explanation of our proposed ML-based decoding schemes, in the following, we use ZF precoding as an example. However, the proposed decoding frameworks are valid for any precoding technique used at the BS.

\subsection{ML framework for the proposed soft decoding scheme}

\begin{figure*}
\centering
  \includegraphics[width=\textwidth]{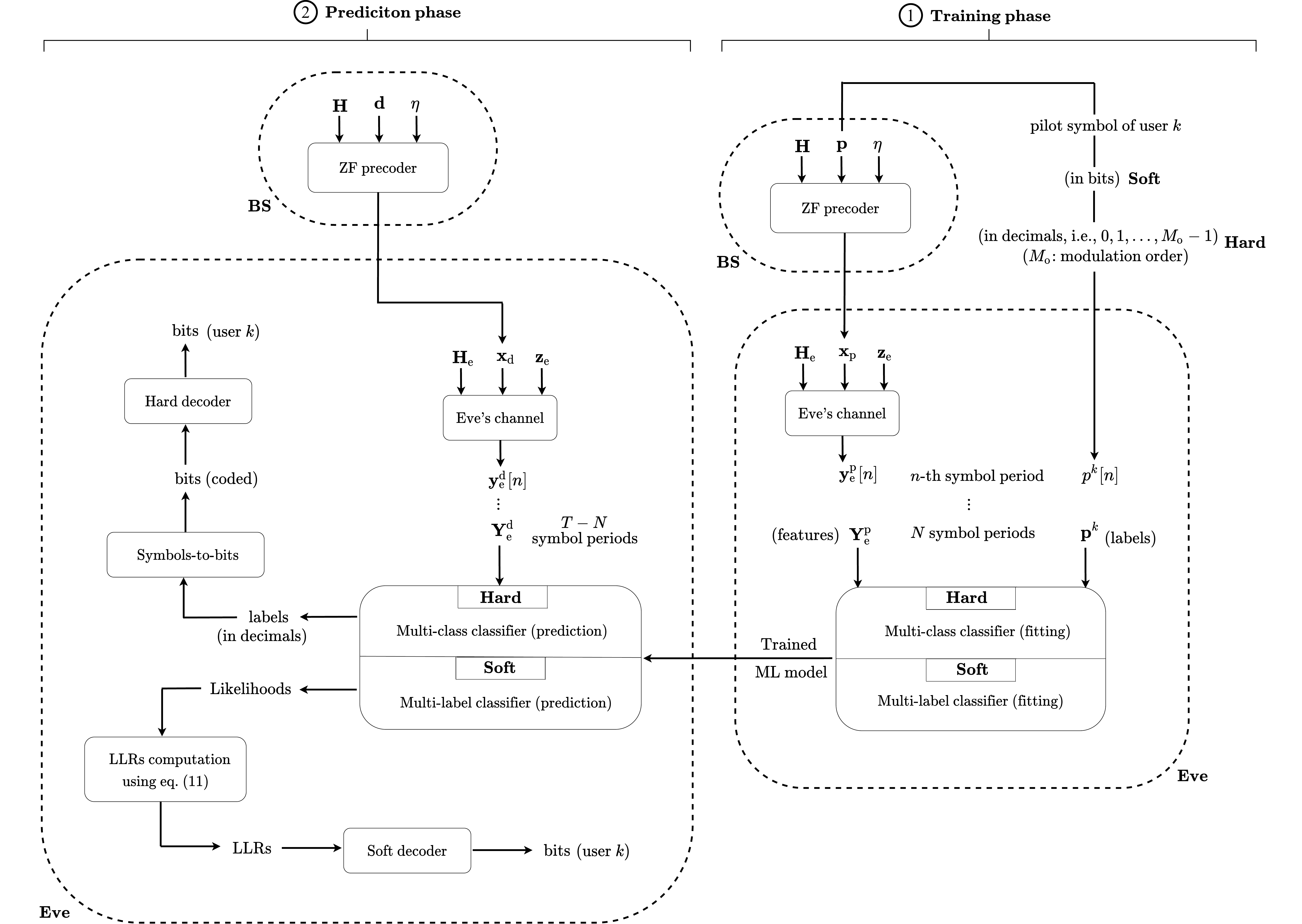}\\ 
  \caption{Training and prediction phases of the ML-based decoding schemes.}
  \label{LLR_algo}
\end{figure*}

As illustrated in Fig.~ \ref{LLR_algo}, the ML framework for soft-decoding encompasses two steps: 1) training phase, where the ML model is trained by using the precoded pilot symbols; 2) prediction phase, where probabilities are estimated and employed to calculate the ${\rm LLRs}$ which are consequently fed to a soft decoder. 

\subsubsection{\textbf{Training phase}}

As pointed out earlier, Eve uses the channel-coded pilot symbols $\mathbf{p} \in\mathbb{R}^{K \times 1}$ as training data, which are pseudo-random sequences for all users. For one symbol period, the overall received pilot signal at Eve's all antennas, $\mathbf{y}^{\rm p}_{{\rm e}}  \in\mathbb{C}^{M \times 1}$, can be written as 

\begin{equation} 
{\mathbf{y}^{\rm p}_{{\rm e}} =  \mathbf{H}_{{\rm e}} \mathbf{x}_{\rm p} + \mathbf{z}_{{\rm e}}},
\end{equation}
\noindent where $\mathbf{x}_{\rm p} \in\mathbb{C}^{N_{\rm t}\times 1}$ is the transmitted precoded pilot signal from the $N_{\rm t}$ BS's transmit antennas.

As depicted in Fig.~ \ref{LLR_algo}, the transmitted signal $\mathbf{x}_{\rm p}$ is depending on all the users' symbols. Thus, the eavesdropping attack can target any user, since pilot symbols are known entities in communication systems. In other words, Eve could target any user individually by retraining the ML model according to the pilot sequences used for each user. As such, Eve would create a mapping between the received signal $\mathbf{y}^{\rm p}_{{\rm e}}$ and the pilot symbols $p^k$ of the targeted user $k$. 

We note that, as the number of antennas at Eve, $M$, increases, the number of received signals at Eve increases accordingly, which often leads to better accuracy. In essence, each antenna at Eve receives a different distorted version of the same transmitted signal $\mathbf{x}_{\rm p}$; the more different copies of $\mathbf{x}_{\rm p}$ received by Eve, the better the ML model performance will be. 

Hence, the training set ${\cal D}$ is the collection of $\{ \mathbf{y}^{\rm p}_{{\rm e}}[n], p^k[n]\}, n \in \{1, \ldots, N\}$, where $\mathbf{y}^{\rm p}_{{\rm e}}[n]$ represents the received pilot signal at Eve during the $n$-th symbol period, while $p^k[n]$ is the corresponding pilot symbol of user $k$. Therefore, the training set ${\cal D}$ can be written in a more compact form as

\begin{equation} 
{\cal D} = \{ \mathbf{Y}^{\rm p}_{{\rm e}}, \mathbf{p}^k \},
\end{equation}
\noindent where $ \mathbf{Y}_{\rm e}^{\rm p} \in\mathbb{C}^{N \times M}$ are the received pilot symbols at Eve during $N$ symbol periods and $\mathbf{p}^k \in\mathbb{C}^{N\times 1}$ are the corresponding transmitted pilot symbols for the $k$-th user. Using ML terminology, $ \mathbf{Y}_{\rm e}^{\rm p}$ represents the features\footnote{We note that the input features in $\mathbf{Y}_{\rm e}^{\rm p}$ are of complex nature and cannot be directly processed by ML algorithms in general. A simple transformation to tackle this incompatibility is to consider the real and imaginary parts of $\mathbf{Y}_{\rm e}^{\rm p}$ as two real numbers. A similar transformation should also be applied to the input features in the validation dataset $\mathbf{Y}_{\rm e}^{\rm d}$, as well. \label{features}} while $\mathbf{p}^k$ represents the labels, where both constitute the training dataset. For non-binary modulation schemes, this ML problem is considered as a multi-label classification (MLC) problem \cite{MLC_BR,MLC_CC} as more than $1$ bit is required to encode the symbols. Namely, MLC is a supervised learning problem where an observation, i.e, a scalar or a vector of features, is associated
with multiple labels. Hence, as depicted in Fig.~ \ref{LLR_algo}, the training dataset is fed to the MLC module which will in turn output a multi-label trained ML model, that will subsequently be used in the prediction phase.  

\subsubsection{\textbf{Prediction phase}}

As depicted in Fig.~ \ref{LLR_algo}, in each symbol period, the BS sends the channel-coded symbols $\mathbf{d}$ to the $K$ users after precoding them using the same scheme employed in the previous phase. The received signals at Eve in each symbol period, $\mathbf{y}^{\rm d}_{{\rm e}} \in\mathbb{C} ^{M \times 1}$, can be written as

\begin{equation} 
{\mathbf{y}^{\rm d}_{{\rm e}} =  \mathbf{H}_{{\rm e}} \mathbf{x}_{\rm d} + \mathbf{z}_{{\rm e}}},
\end{equation} 

\noindent where $\mathbf{x}_{\rm d} \in\mathbb{C}^{N_{\rm t} \times 1}$ represents the transmitted precoded data from the $N_{\rm t}$ transmit antennas, intended for all the users during one symbol period. 
If we assume that there are $T$ symbols in one coherence time, $\mathbf{Y}^{\rm d}_{{\rm e}} \in\mathbb{C}^{(T-N) \times M}$ represents the collection of all received signals at Eve during one coherence time of the transmitted $(T-N)$ data symbols. In ML nomenclature, $\mathbf{Y}^{\rm d}_{{\rm e}}$ is commonly being referred to as the test/evaluation dataset.

In principle, the goal of classification is to predict labels. In this context, however, we are not interested in the labels (hard outputs) but rather in the corresponding probabilities (soft outputs) to be used subsequently for LLR computation. Before tackling the computation of these probabilities, we first provide an overview of ${\rm LLRs}$ computation based on probabilities. We start by recalling some fundamental definitions in the context of ${\rm LLR}$ computation in binary detection \cite{Galager_OCW}. 

Let $U$ be a binary random variable (RV), acting as the correct hypothesis, with possible values $\{ a_0,a_1 \}$ and a priori probabilities $p_0$ and $p_1$. Let $V$ be an RV with conditional probability density $f_{V|U}(v|a_m)$ that is finite and non-zero for all $v \in \mathbb{R}$ and $m \in \{0,1\}$. In our context, $V$ models the received signal at Eve at a fixed time instant. We note that the conditional densities $f_{V|U}(v|a_m), m \in \{0,1\}$, are called \textit{likelihoods}. The marginal density of $V$ is given by $f_{V}(v) = p_0 f_{V|U}(v|a_0) + p_1 f_{V|U}(v|a_1)$. Hence, the a posteriori probability of $U$ can be expressed as 
 
  \begin{equation} \label{APost_Prob}
f_{U|V}(a_m|v) = \frac{p_m f_{V|U}(v|a_m)}{f_{V}(v)},
\end{equation}

 \noindent where $m \in \{0, 1\}$. To maximize the probability of correct detection, the maximum a posteriori (MAP) rule can be written as 
 
  \begin{equation} \label{MAP}
\frac{p_0 f_{V|U}(v|a_0)}{f_{V}(v)}   \mathop{\gtreqless}\limits_{\tilde{U} = a_1}^{\tilde{U} = a_0}  \frac{p_1 f_{V|U}(v|a_1)}{f_{V}(v)},
\end{equation}
\noindent where $\Tilde{U}$ denotes the decision on the RV $U$. Rearranging (\ref{MAP}) and canceling $f_{V}(v)$, we obtain the \emph{likelihood ratio}

 \begin{equation} \label{Lambda}
\Lambda(v) = \frac{ f_{V|U}(v|a_0)}{f_{V|U}(v|a_1)}   \mathop{\gtreqless}\limits_{\tilde{U} = a_1}^{\tilde{U} = a_0}  \frac{p_1}{p_0} ,
\end{equation}

\noindent where the quantity $ \frac{p_1}{p_0}$ is called the \textit{threshold} and depends only on the a priori probabilities. Hence, the log-likelihood ratio ${\rm LLR}(v)$ can be expressed as follows:

 \begin{equation} \label{LLR}
{\rm LLR}(v) = \ln \bigg[ \frac{ f_{V|U}(v|a_0)}{f_{V|U}(v|a_1)} \bigg] .
\end{equation}

As depicted in Fig.~\ref{LLR_algo}, to obtain the likelihoods in eq. (\ref{LLR}), we feed the test dataset to the MLC module. A common and efficient implementation of predicting these probabilities is the Platt scalling approach in \cite{Platt99}. This method is used to transform the uncalibrated outputs of the classification module into probabilities. Platt scaling works by fitting a logistic regression model to the classifier's scores. The probabilities $f_{V|U}(v|a_m)$ according to the Platt scaling algorithm can be computed as 

\begin{equation} \label{Platt)}
f_{V|U}(v|a_m) = \frac{1}{1+\exp(A f(a_m) + B)},
\end{equation}

\noindent where $f(a_m)$ is the classifier score and scalars $A$ and $B$ are the sigmoid parameters \cite{Platt99} learned by the algorithm, which are calculated using a cross-entropy loss function and an internal threefold cross-validation to prevent overfitting. 

Once the likelihoods $f_{V|U}(v|a_m)$ are obtained, the ${\rm LLRs}$ can be computed using eq. (\ref{LLR}), after which Eve can simply feed the computed ${\rm LLRs}$ to the soft decoder to obtain the transmitted message to user $k$.
 
\subsection{ML framework for the proposed hard decoding scheme}

Similar to the soft decoding scheme, the ML-based hard decoding scheme comprises of two phases: 1) training phase, where the ML model is trained by using SLP precoded pilot symbols; 2) prediction phase, where the module directly predicts the transmitted symbols, which are in turn mapped into bits to finally be fed to a hard decoder.

\subsubsection{\textbf{Training phase}}
As depicted in Fig.~\ref{LLR_algo}, this is almost the same as the training phase of the soft decoding scheme, with the exception that pilot symbols of user $k$ are represented as decimals instead of bits. Therefore, this problem is a single-label multi-class classification (MCC) problem. For a modulation order of $M_{\rm o}$, the label space is  $\{ 0,1, \ldots, M_{\rm o}-1 \}$ with $M_{\rm o}$ total classes. The output of this phase is a trained ML model, that will be used next in the prediction phase. 

\subsubsection{\textbf{Prediction phase}}
Given Fig.~ \ref{LLR_algo}, this step is also mostly equivalent to the previously detailed prediction phase of the soft decoding scheme. However, it differs in the classification module used and the prediction method. Similar to the training phase, the prediction module employs MCC to obtain the estimated data symbols in the form of decimals, i.e., the same nature of the labels used in the training phase. Once the data symbols are predicted, they will be first mapped into bits and then fed to a hard decoder for decoding. 

\section{Countermeasure: Physical-layer Security }\label{CM}
In this section, we propose novel security-enhanced SLP schemes that yield high BER at Eve. Similar to \cite{Abdu_CNS} and \cite{Abdu_OJ_COMs}, the idea is to design the transmitted signal $\mathbf{x}_{\rm d}$ to have constructive interference at the intended users, while at the same time, increasing the BER at Eve. We note that the CSI to Eve is available at the BS. This assumption is reasonable when Eve is a legitimate user attempting to eavesdrop other users \cite{Masouros}. This assumption gives Eve the advantage to access the control signaling of the BS and obtain the modulation and coding parameters used, which further improves its detection performance.

\subsection{PLS random scheme}

\begin{figure}[t]
\centering
  \includegraphics[width=2.7in]{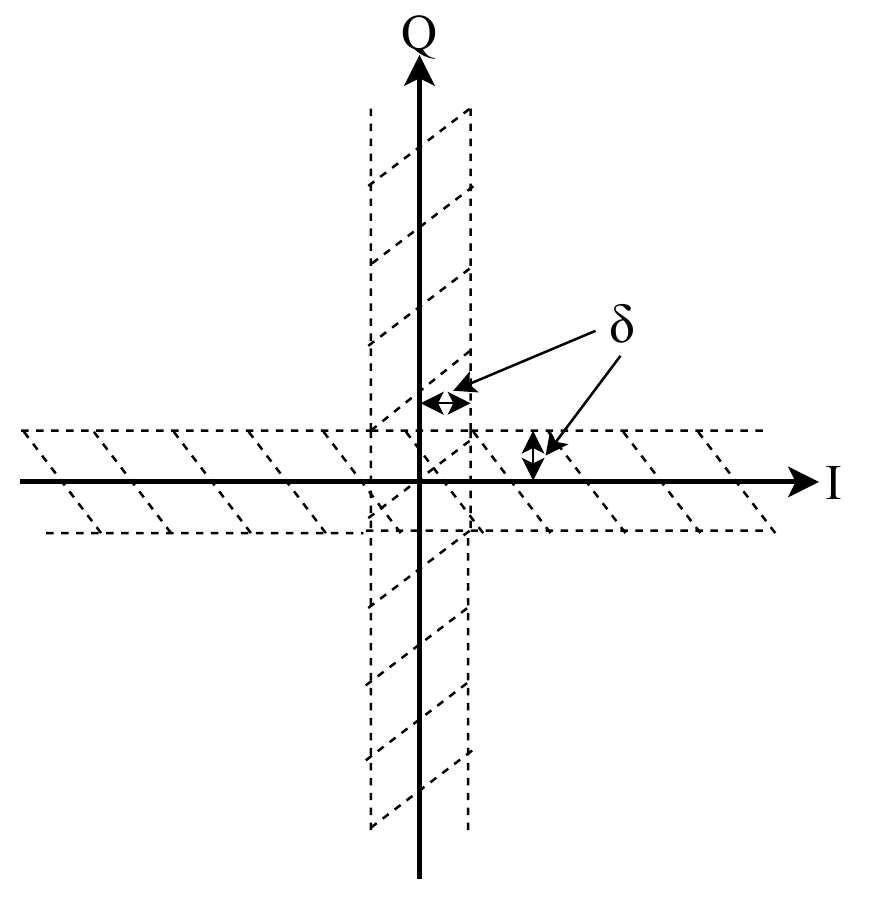}\\ 
  \caption{Boundary region for Eve's received signal in the case of PLS random scheme.}
  \label{Boundary_scheme}
\end{figure}

Inspired from the boundary scheme presented in \cite[\S IV.B eq. (11)]{Abdu_OJ_COMs}, which forces Eve's received signal to lie very close to the boundary of the detection region, we propose to embed randomness in Eve's received signal by randomly selecting the boundary region, either horizontal or vertical, as depicted in Fig.~\ref{Boundary_scheme}, at each symbol period to further obstruct the learning process at Eve. We refer to this scheme as ``PLS random scheme". The transmission signals design for this scheme can be formulated as   

\begingroup
  \small
\begin{IEEEeqnarray}{rCl} 
 \noindent \mathbf{x}_{\rm d} (\mathbf{d}, \mathbf{H}, \mathbf{h}_{\rm e}, \boldsymbol{\gamma}, {\delta}, {\mathbf{b}})  &{=}& 
\displaystyle \arg \min_{\mathbf{x}}  {\lVert \mathbf{x}\lVert^2 } \label{slp:main}  \\
&& \text{subject to} \nonumber\\ 
&& {\rm Re}\{\mathbf{h}_k \mathbf{x} \} \trianglelefteq \sigma_z {\sqrt{\gamma_k}} {\rm Re}\{d_k\}, \; \forall k \label{slp:constci1}\\
&& {\rm Im}\{\mathbf{h}_k \mathbf{x} \} \trianglelefteq \sigma_z {\sqrt{\gamma_k}} {\rm Im}\{d_k\}, \; \forall k \label{slp:constci2} \\ 
&& b_i {\rm Re}\{\mathbf{h}_{\rm e}^i \mathbf{x} \} {+} {(1{-}b_i)}{\rm Im}\{\mathbf{h}_{\rm e}^i \mathbf{x} \} {\lesseqgtr} {\delta}, {\forall}{i}  \label{slp:consteve1}
 \end{IEEEeqnarray}
\endgroup

\noindent where $\mathbf{h}_k \mathbf{x}$ is the $k$-th user's noiseless received signal, $\mathbf{h}_{\rm e}^i \mathbf{x}$ is the noiseless received signal at Eve's $i$-th antenna, $b_i \in \{0,1\}$ is the realization of a binary RV that represents the boundary region to use for Eve's $i$-th antenna, $\mathbf{b}$ is the vector collecting the $b_i$ realizations of the $M$ antennas at Eve, $\trianglelefteq$ defines the correct\footnote{For further detailed information, the reader should refer to~\cite{Danilo_TSP_18}.} detection region,  and $\delta > 0$ is the distance parameter controlling the width of the boundary region. This problem is convex and can be solved efficiently using standard optimization toolboxes such as CVX \cite{Boyd}. 

To illustrate the PLS random schemes, we decompose the non-convex strapped region in Fig.~\ref{Boundary_scheme} into two convex regions: the vertical part when $b_i = 1$ and the horizontal one when $b_i = 0$, corresponding to real and imaginary parts of constraint (\ref{slp:consteve1}). And at each symbol period and for each antenna at Eve, we \textit{randomly} choose between the two regions, so that on average, Eve's received signals would lie evenly on both regions, assuming the symbol distribution is equiprobable.

\subsection{PLS Eve-min-power scheme}
The proposed PLS scheme is designed to have less complexity compared to the previous scheme, i.e., no Eve-related constraints, while still keeping a high BER at Eve. This scheme can be formulated as follows:

\begingroup
  \small
\begin{IEEEeqnarray}{rCl} \label{SLP_PLS}
\noindent \mathbf{x}_{\rm d} (\mathbf{d}, \mathbf{H}, \mathbf{H}_{\rm e}, {\boldsymbol{\gamma}}) &{=}& 
\displaystyle \arg \min_{\mathbf{x}_{\rm d}}  {||\mathbf{x}_{\rm d}|| + ||\mathbf{H}_{\rm e} \mathbf{x}_{\rm d}|| } \label{slp1:main} \\
&& \text{subject to} \nonumber\\ 
&& {\rm Re}\{{\mathbf{h}_k} {\mathbf{x}_{\rm d}} \} {\trianglelefteq} \sigma_z \sqrt{\gamma_k} {\rm Re}\{d_k\}, \; \forall k  \label{slp1:constci1}\\
&& {\rm Im}\{{\mathbf{h}_k} {\mathbf{x}_{\rm d}} \} {\trianglelefteq} \sigma_z \sqrt{\gamma_k} {\rm Im}\{d_k\}, \; \forall k \label{slp1:constci2} .
\end{IEEEeqnarray}
\endgroup

The PLS Eve-min-power scheme minimizes the sum of the transmit power and the noiseless received power at Eve, as shown in (\ref{slp1:main}), while guaranteeing a certain target SINR at the intended users through constructive interference constraints (\ref{slp1:constci1}) and (\ref{slp1:constci2}). Similarly, this problem is convex and can be solved efficiently using standard optimization toolboxes.

\section{Numerical results}\label{NR}

To make this section more comprehensive, we split it into three parts: 1) Parameters, metrics, and benchmarks where we defined the simulations' setting, 2) selection of ML algorithms for Eve attack in which we experiment with several algorithms and select the most performing, and 3) comparisons and insights to assess the performance of our proposed schemes in terms of security, power consumption, and complexity. 

\subsection{Parameters, metrics, and benchmarks}

As a benchmark to the proposed SLP-based PLS schemes, we employed the CISPM \cite{Maha_TSP_15} and ZF precoding \cite{ZF_Andrea} schemes. We note that in the following simulations, for a fair comparison, we set $\eta = \gamma_k$ such that all the examined schemes have the same transmit power.  

Regarding the metrics used to evaluate the different schemes, we use the BER at Eve to assess the security offered by a particular decoding scheme for a given precoding design. The lower the BER at Eve, the lower the security and vice versa. In a similar way, we also evaluate the frame-error rate (FER) at Eve since we have channel coding in the system, which is defined as the ratio of frames in error (one altered bit suffices to make the entire frame erroneous) to the total number of frames received. We also evaluate the BER/FER at the intended user to examine the impact of using the PLS schemes on the intended user's performance. Finally, we define the total transmit power by the BS antennas as $P_{{\rm tot}} = \Vert {\bf x}_{\rm d}\Vert^2$. In the simulations, we take the average of the above quantity over a large number of symbol slots to obtain the frame-level total transmit power, which is then averaged over a large number of channel realizations.

In the following simulations, we use QPSK modulation. For the PLS random scheme, we set $\delta = \sfrac{\sigma_z^2}{10}$ to make sure that the noise will push Eve's received signal outside the boundary region, i.e., to cause higher error rates at Eve. For simplicity, we consider unitary noise variance $\sigma_z^2$. As for the channel coding part, we use convolutional coding \cite{Coding_Matlab} and Viterbi decoding \cite{Viterbi} with the parameters in Table~\ref{Tab_FEC}. We note that low coding rates are chosen in order to consider a worst case eavesdropping scenario, where Eve can take advantage of the redundancy to correct as much errors as possible. 

\begin{table}[t!]
  \begin{center}
    
    \label{tab:table1}
    \begin{tabular}{|l|c|r}
       \hline
      \textbf{Parameters} & \textbf{Values} \\  \hline
      Code rates & \sfrac{1}{3}, \sfrac{1}{4} \\ \hline
      Constraint length & 7 \\ \hline
      Frame Size & 150  \\ \hline
      Number of frames & 100  \\ \hline
      Trace-back length & 96 \\ \hline
     Decoder decision technique & Hard, Soft\\ \hline  
    \end{tabular}
  \end{center}
  \caption{Channel coding parameters used for the simulations}
  \label{Tab_FEC}
\end{table}

\subsection{Selection of ML algorithms for the Eve attack}

For the MLC modules used for the ML-based soft decoding scheme, in this simulation, we adopt problem transformation methods that remodel our MLC problem into single-label problem(s). Since our labels are bits, the MLC problem will be decomposed into $k$ binary classifiers, where $k = \log_2 M_o$ is the number of bits constructing each symbol. 

Herein, we use two transformation methods, binary relevance (BR) \cite{MLC_BR} and classifier chain (CC) \cite{MLC_CC}. BR is the most simple and efficient method to solve MLC problems, which trains the $k$ binary classifiers independently; its only drawback is that it does not consider labels correlation. CC, however, takes into account the correlation between labels by using the outputs of the previously trained classifiers as features for the subsequent ones in the chain, except for the first classifier. We refer to these soft-decoding implementation by ``Soft - BR" and ``Soft - CC" accordingly. 

Concerning the MCC module used for the ML-based hard decoding scheme, it does not require any transformation or specific approach. It can be solved using any classifier. We refer to this scheme subsequently by ``Hard". It is worth mentioning that the ML-based decoding schemes apply only to Eve, whereas the intended users employ conventional (not ML-based) soft and hard decoding techniques. 

To make Eve as sophisticated as possible, we experiment with several state-of-the-art classifiers and choose the one with the best performance. In Table~\ref{tab:1}, we compare the prediction accuracy of the proposed soft\footnote{In the table, we did not show results for Soft - BR to avoid redundancy, as its results were almost the same as Soft - CC.} and hard decoding schemes, considering ZF and CISPM precoding as well as the proposed PLS precoding schemes. The parameters used for this simulation are: $N_{\rm t} = 15$, $K = 6$, $M = 9$, and $\eta = \gamma_k = 6~\mathrm{dB}$. We note that these results represent the averaged results over $100$ different channel realizations. We also note that this accuracy applies before channel decoding, i.e., by comparing the ML predicted labels to the actual coded transmitted symbols to user $k$. 

\begin{table*}[t]
  \centering
   \begin{tabular}{|p{2cm}|c|c|c|c|c|c|c|c|}
    \hline
    \multirow{2}{5cm}{\textbf{Classifiers}} &
    \multicolumn{2}{c|}{\textbf{ZF}} &
    \multicolumn{2}{c|}{\textbf{CISPM}} &
    \multicolumn{2}{c|}{\textbf{PLS random}} & \multicolumn{2}{c|}{\textbf{PLS Eve-min-power}}\\
    \cline{2-9}
    &  \textbf{Soft - CC} & \textbf{Hard}  & \textbf{Soft - CC} & \textbf{Hard} & \textbf{Soft - CC} & \textbf{Hard}  & \textbf{Soft - CC} & \textbf{Hard}\\
     \hline
    \text{Gaussian\_NB} & 0.8338& 0.8378& 0.8271& 0.8298& 0.5431& 0.5429& 0.5708& 0.5712\\
    \hline
    \text{Log\_Reg} \cellcolor{blue!25} & 0.9375& 0.9374& 0.8935& 0.8929& 0.5427& 0.5433& 0.5732& 0.5732\\ \hline
    SVM & 0.9370& 0.9361& 0.8925& 0.8931& 0.5429& 0.5426& 0.5697& 0.5718\\ \hline
    \text{R\_Forest} & 0.8831& 0.8798& 0.8538& 0.8491& 0.5382& 0.5354& 0.5669& 0.5650  \\ \hline
    KNN & 0.9047& 0.8994& 0.8623& 0.8570& 0.5265& 0.5245& 0.5474& 0.5454   \\ \hline
    \text{Decision\_Tree} & 0.8101& 0.7951& 0.7738& 0.7571& 0.5178& 0.5207& 0.5372& 0.5374   \\ \hline
    \text{Extra\_Trees} & 0.9017& 0.8970& 0.8669& 0.8644& 0.5387& 0.5377& 0.5670& 0.5679   \\ \hline
    \text{LightGBM} & 0.8998& 0.8958& 0.8655& 0.8611& 0.5359& 0.5356& 0.5645& 0.5629   \\ \hline
    \text{XGB} & 0.8983& 0.8946& 0.8633& 0.8609& 0.5326& 0.5349& 0.5598& 0.5605 \\ \hline
  \end{tabular}
  \caption{Prediction accuracy of our proposed ML-based decoding schemes with several classifiers when using ZF, CIPSM, PLS random, and PLS Eve-min-power precoding at the BS. }
  \label{tab:1}
\end{table*}

As observed in Table~\ref{tab:1}, the logistic regression classifier achieves the highest prediction accuracy among all the precoding and decoding schemes. Therefore, in the following simulations, we use this classifier in our proposed ML-based decoding schemes. 

\subsection{Comparison and insights}
We note that a BER value of $0.5$ indicates full confusion. Regarding the target FER values at the intended users, it varies depending on the application scenario. For instance, enhanced mobile broadband (eMBB) in 5G requires an FER on the order of $10^{-3}$ while massive machine-type communications (mMTC) require only $10^{-1}$ \cite{5G_FER}. In this section, we will validate the eavesdropping attacks by showing the FER at Eve to be in the order of the intended users' FER values. 

\begin{figure}[!htb]
\minipage{0.51\textwidth}
  \includegraphics[width=\linewidth]{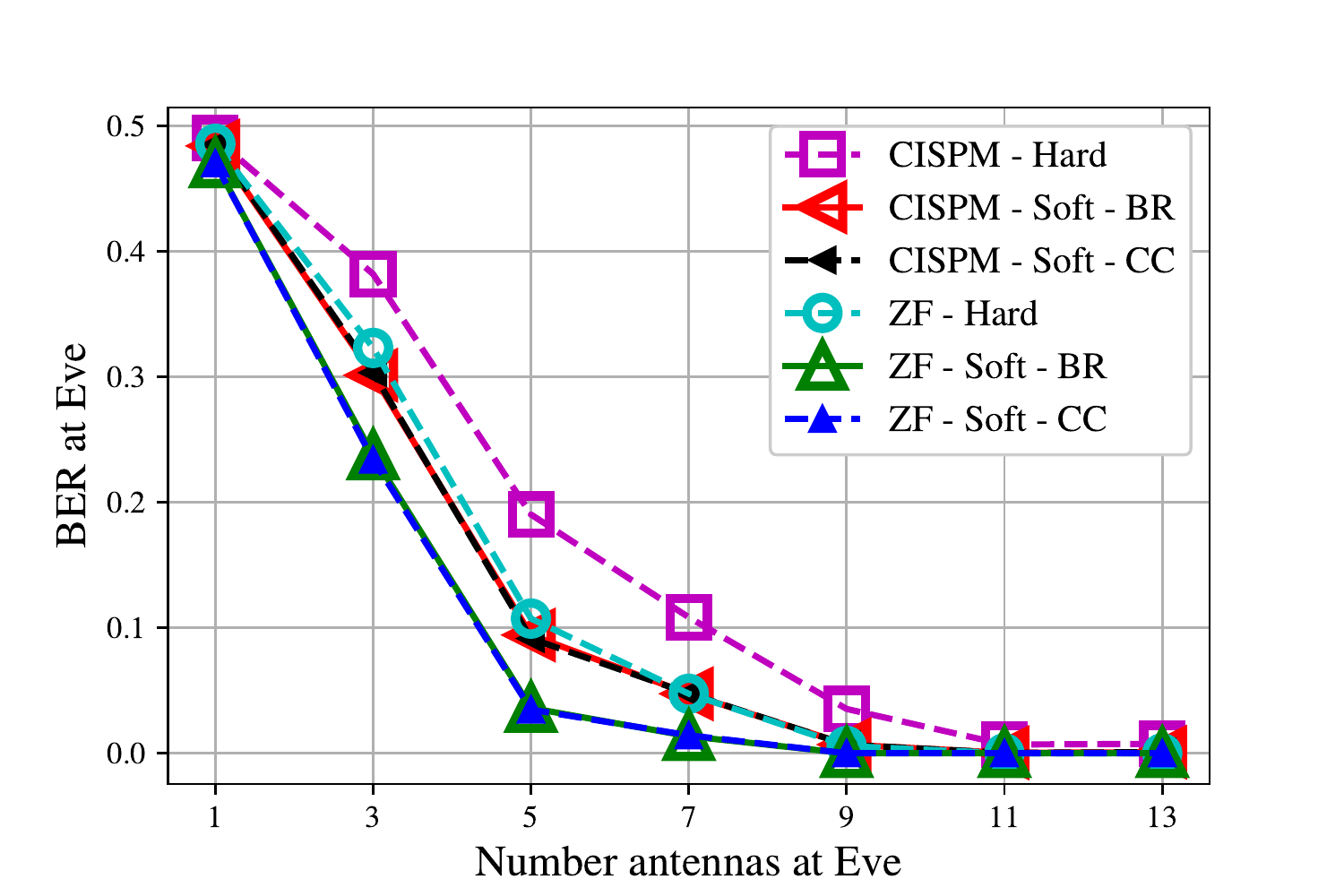}
  \caption*{(a) ZF and CISPM schemes}\label{fig:awesome_image1}
\endminipage\hfill
\minipage{0.51\textwidth}
  \includegraphics[width=\linewidth]{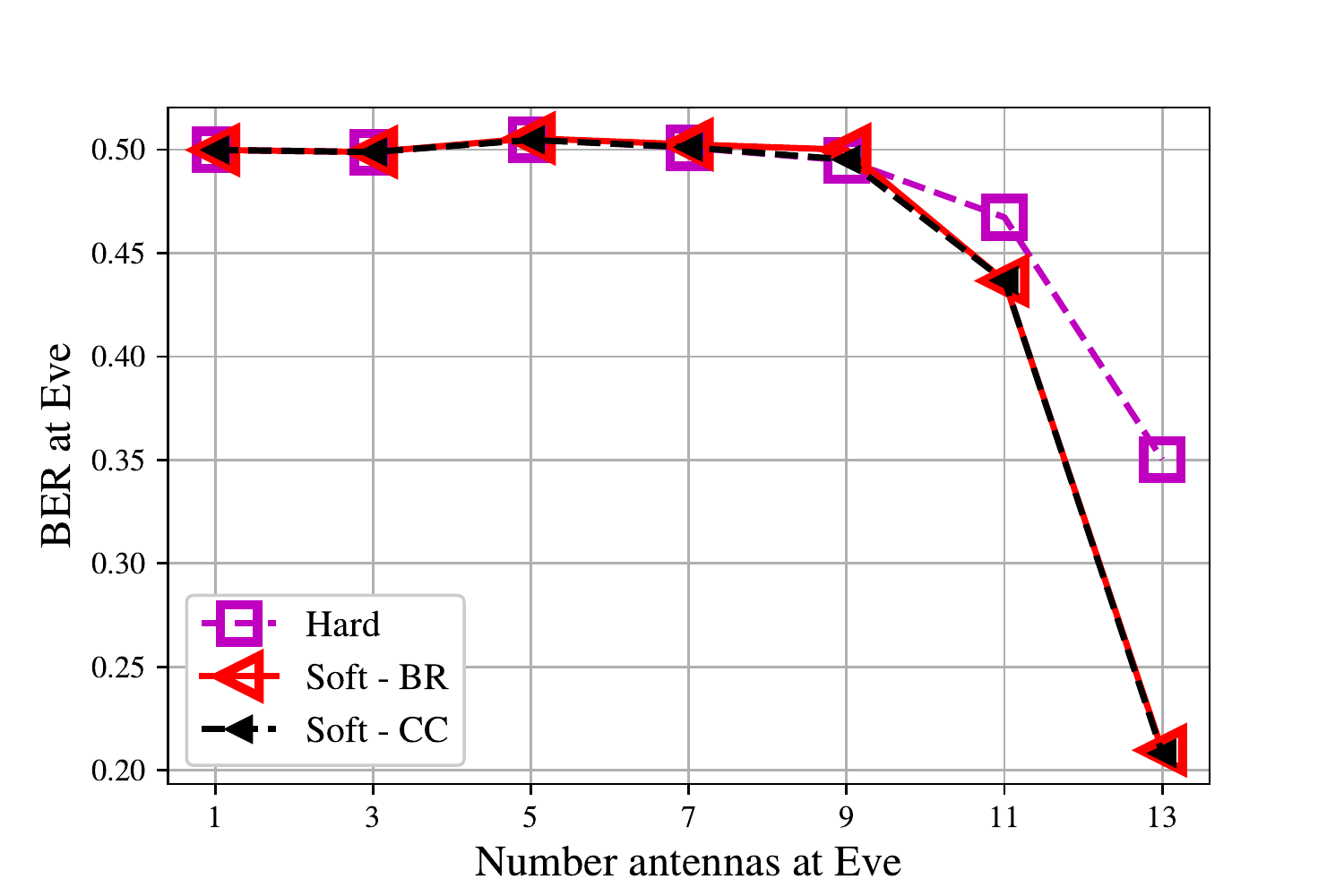}
  \caption*{(b) PLS random scheme}\label{fig:awesome_image2}
\endminipage\hfill
\minipage{0.51\textwidth}%
  \includegraphics[width=\linewidth]{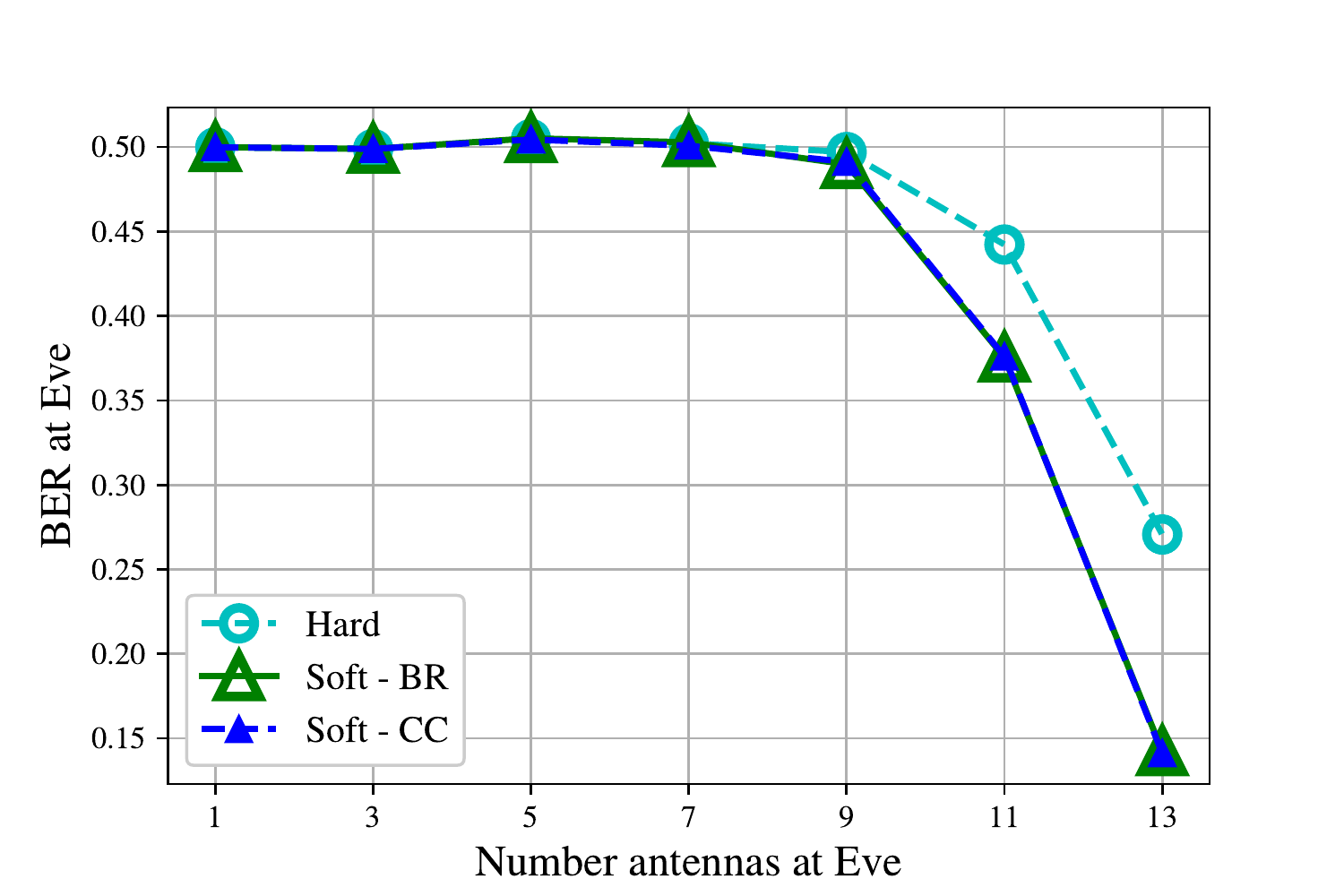}
  \caption*{(c) PLS Eve-min-power scheme}\label{fig:awesome_image3}
\endminipage
\caption{QPSK-coded BER at Eve vs. number of antennas at Eve, with $r=\sfrac{1}{3}$, $N_{\rm t} = 15$, $K = 6$, and $\eta = \gamma_k = 6~\mathrm{dB}$.}
\label{BER_vs_nAnt13}
\end{figure}

Fig.~\ref{BER_vs_nAnt13} depicts the coded BER at Eve as a function of its number of antennas $M$. We compare the proposed hard and soft decoding schemes. The parameters used in the simulation are: $r=\sfrac{1}{3}$, $N_{\rm t} = 15$, $K = 6$, and $\eta = \gamma_k = 6~\mathrm{dB}$. Fig.~\ref{BER_vs_nAnt13}(a) represents the non-secure precoding schemes, ZF and CISPM. For our hard and soft decoding schemes, we observe that the more antennas at Eve, the lower is the BER, i.e., the more antennas at Eve, the higher the prediction accuracy (more versions of the same signal, hence more features used for training and prediction), leading to lower BER. In addition, our Soft technique outperforms the Hard one, where Soft - BR and Soft - CC are equivalent. Moreover, with $9$ antennas at Eve, the BER at Eve is so low to the point that it could be compared to an intended user's decoding performance, leading to a big vulnerability in systems that use ZF and CISPM precoding. In addition, as expected, CISPM is more secure than ZF as predicted in Sec.~\ref{ML-A}. As for the case of the PLS random scheme in Fig.~\ref{BER_vs_nAnt13}(b), we observe the same pattern as in Fig.~\ref{BER_vs_nAnt13}(a), the more antennas at Eve, the lower is the BER, with Soft decoding outperforming the Hard one. However, when $M$ values are lower or equal to $9$, the BER at Eve is at $0.5$, indicating total equivocation. In fact, even for higher values than $9$, the BER at Eve is still very high compared to ZF and CISPM schemes, i.e., PLS random scheme is offering a significant security gain when compared to the latter ones. Similarly, for the PLS Eve-min-power scheme in Fig.~\ref{BER_vs_nAnt13}(c), we observe the same behavior as for PLS random scheme, with the PLS Eve-min-power scheme BER going lower than PLS random, i.e., PLS Eve-min-power is less secure than PLS random. However, when compared to non-secure precoding schemes, ZF and CISPM, PLS Eve-min-power still offers a drastic security gain.

\begin{figure}[!htb]
\centerline{\minipage{0.51\textwidth}
  \includegraphics[width=\linewidth]{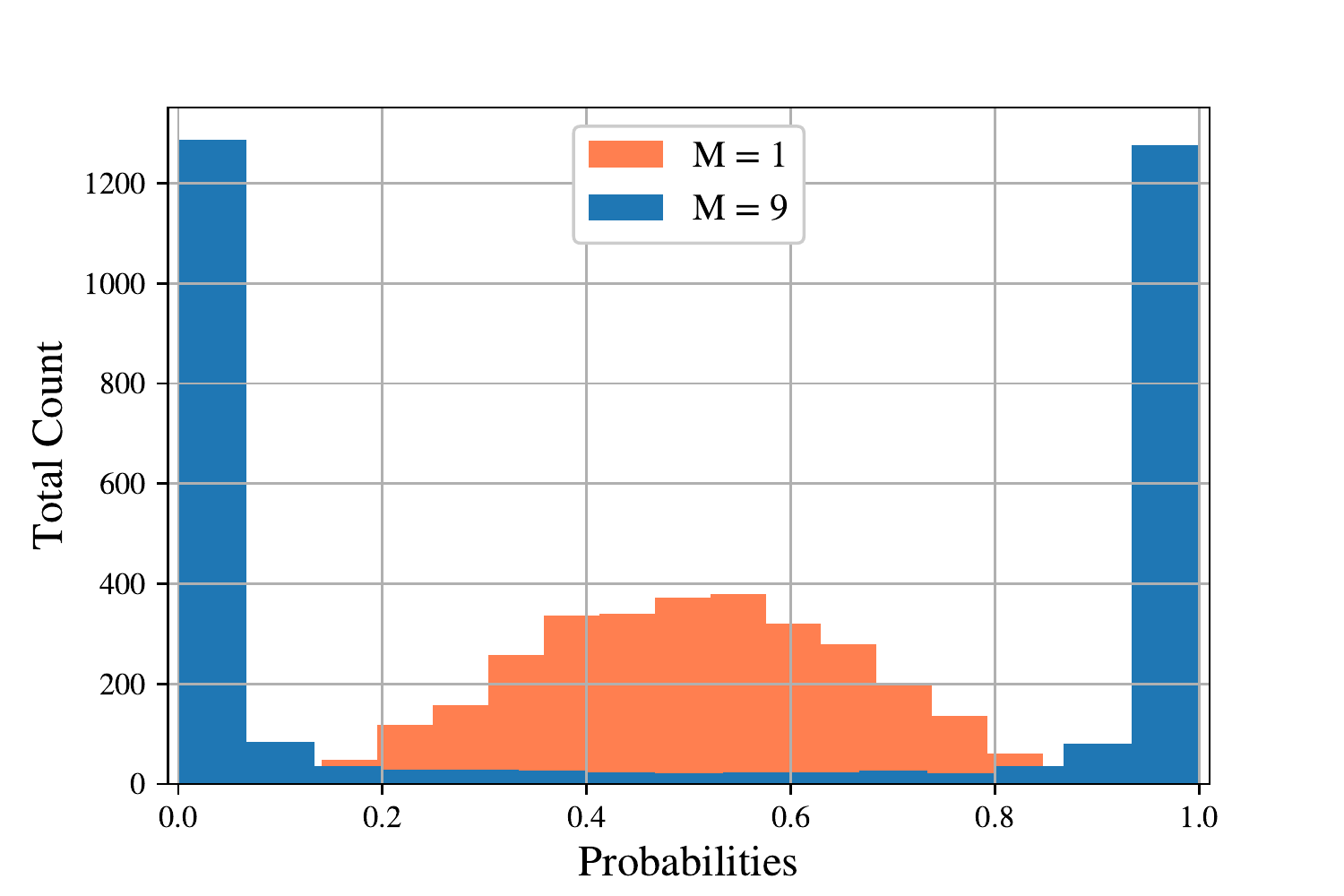}
  \caption*{(a) Probabilities}\label{fig:awesome_image1}
\endminipage}\hfill
\centerline{\minipage{0.51\textwidth}
  \includegraphics[width=\linewidth]{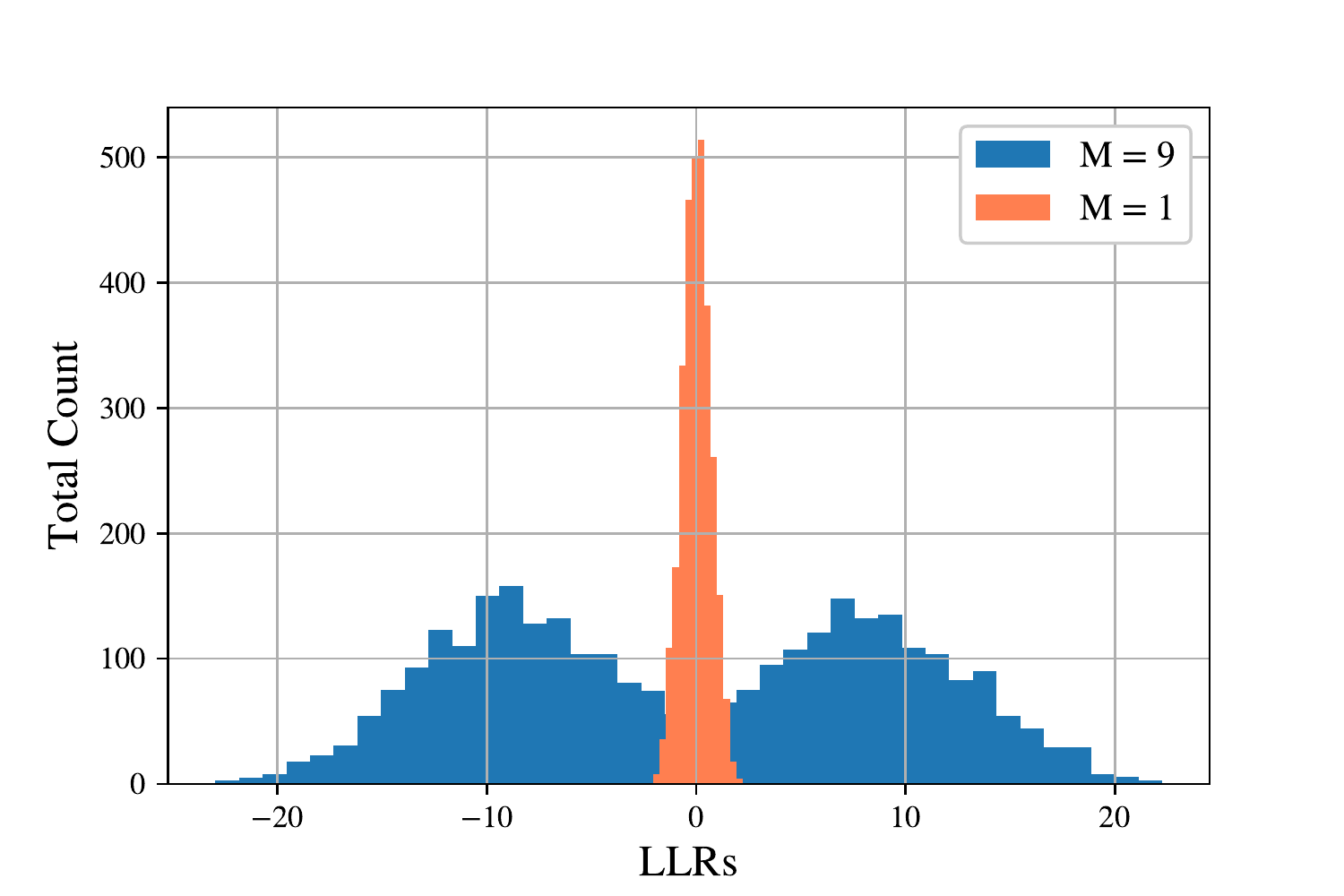}
  \caption*{(b) ${\rm LLRs}$ }\label{fig:awesome_image2}
\endminipage}\hfill
\caption{Histogram of the probability distribution of the classes and ${\rm LLRs}$ for Soft - CC decoding scheme with $r = \sfrac{1}{3}$ and $M \in \{1,9 \}$.}
\label{Post_prob_dist}
\end{figure}

Fig.~\ref{Post_prob_dist}(a) depicts the distribution of the estimated  probabilities and their corresponding ${\rm LLRs}$ for Soft - CC decoding scheme. We note that for this particular plot, we used $1000$ symbols to obtain a smooth histogram. We recall that a probability of $0.5$ indicates that the predictor is not sure whether the predicted bit should be $0$ or $1$; a value close to $1$ means the predictor is very sure that it is a $1$, whereas a probability close to $0$ indicates the opposite, i.e., it is very sure that it is not a $1$. For the case of $M = 1$, we observe that most of the probabilities are close to $0.5$, which demonstrates a poor prediction accuracy. However, when Eve uses $M = 9$ antennas, we observe that most probabilities are at the edges, i.e., either close to $0$ or close to $1$, thus yielding a high prediction accuracy. Using a higher number of antennas entails more features that can be employed in both training and prediction phases, therefore leading to higher accuracy when estimating the likelihoods. Concerning the ${\rm LLRs}$ distribution in Fig.~\ref{Post_prob_dist}(b), as expected, the ${\rm LLRs}$ values for the case of $9$ antennas at Eve are distributed mostly away from $0$, indicating high quality ${\rm LLRs}$, as their corresponding probabilities are mostly different than $0.5$. However, when Eve uses only $1$ receive antenna, the ${\rm LLRs}$ are close to $0$, implying poor quality ${\rm LLRs}$. Therefore, from Figs.~\ref{BER_vs_nAnt13} and \ref{Post_prob_dist}, we conclude that higher number of antennas at Eve leads to higher prediction accuracy, and therefore lower BER.  

\begin{figure}[!htb]
\minipage{0.51\textwidth}
  \includegraphics[width=\linewidth]{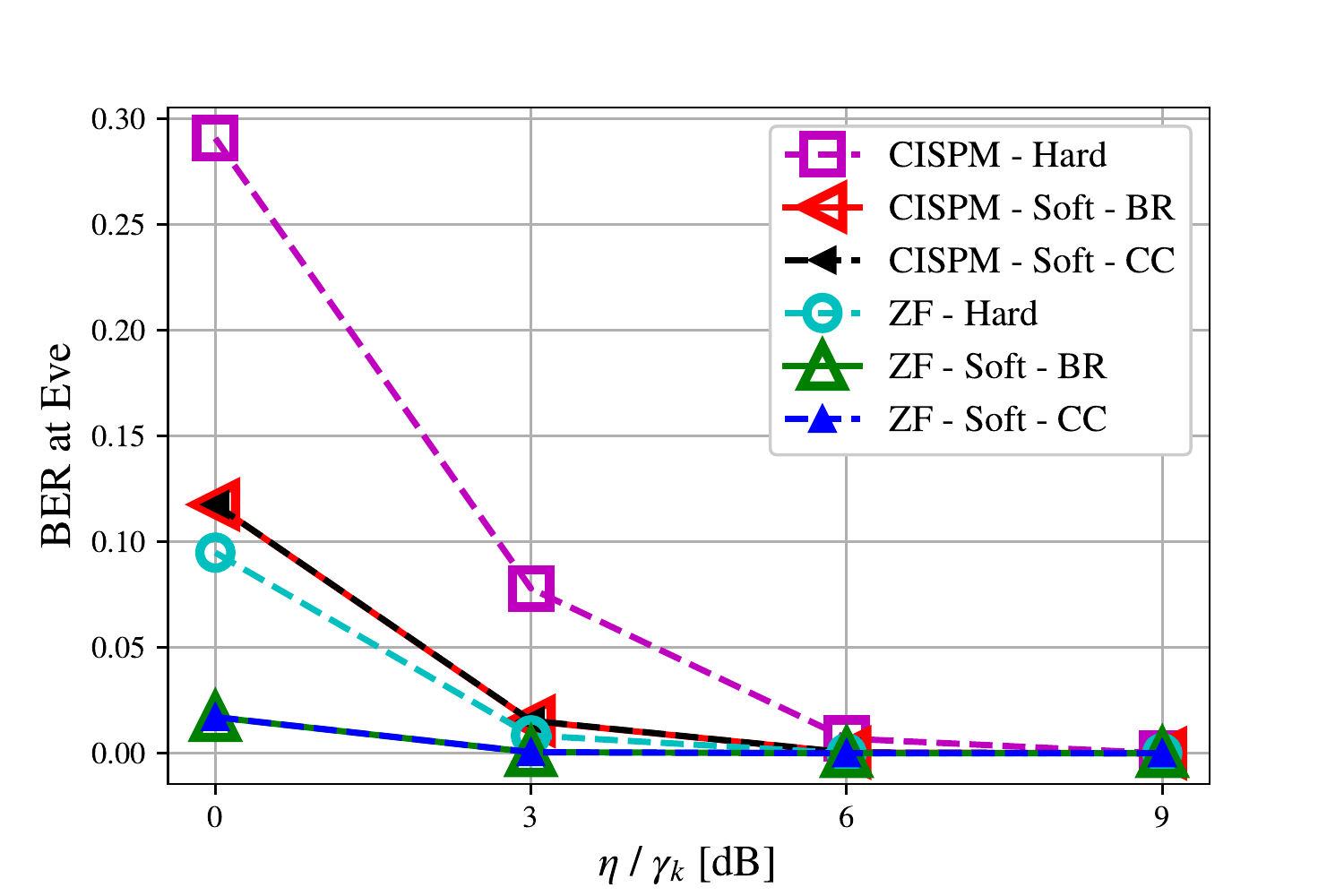}
  \caption*{(a) ZF and CISPM schemes}\label{fig:awesome_image1}
\endminipage\hfill
\minipage{0.51\textwidth}
  \includegraphics[width=\linewidth]{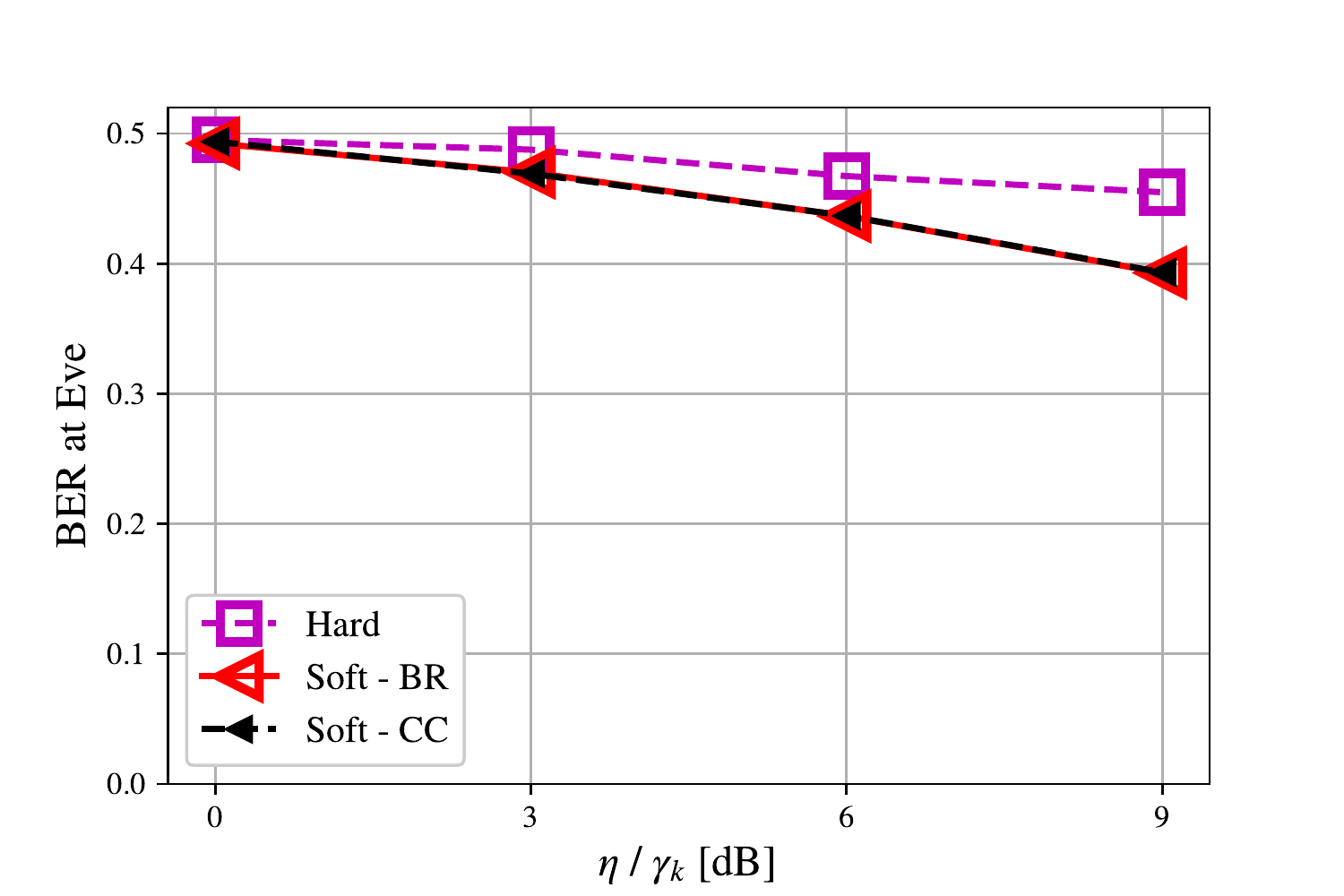}
  \caption*{(b) PLS random scheme}\label{fig:awesome_image2}
\endminipage\hfill
\minipage{0.51\textwidth}%
  \includegraphics[width=\linewidth]{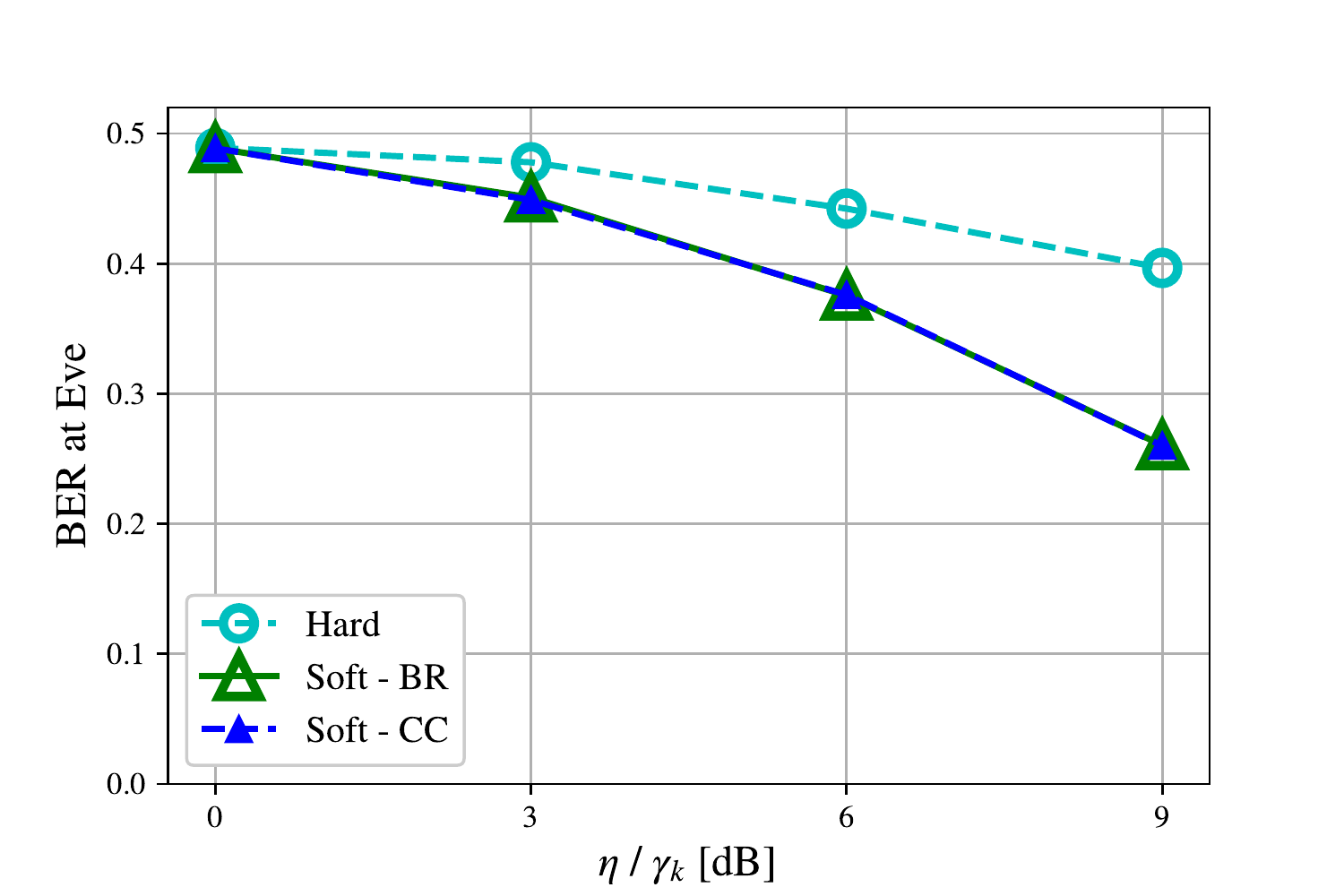}
  \caption*{(c) PLS Eve-min-power scheme}\label{fig:awesome_image3}
\endminipage
\caption{QPSK-coded BER at Eve vs. $\eta/\gamma_k$ $\mathrm{[dB]}$, with $r=\sfrac{1}{3}$, $N_{\rm t} = 15$, $K = 6$, and $M = 11$.}
\label{BER_vs_tSINR_13_3Ant}
\end{figure}

Fig.~\ref{BER_vs_tSINR_13_3Ant} depicts the coded BER at Eve as a function of $\eta/\gamma_k$ $\mathrm{[dB]}$, which we set to the same value for all users for simplicity. The parameters used in the simulation setup are: $r=\sfrac{1}{3}$, $N_{\rm t} = 15$, $K = 6$, and $M = 11$. Concerning ZF and CISPM schemes in Fig.~\ref{BER_vs_tSINR_13_3Ant}(a), with the proposed soft and hard decoding approaches we notice that the higher the values of $\eta/\gamma_k$, the lower the BER. Particularly, higher $\eta/\gamma_k$ values leads to higher transmit power, which cause higher received power at Eve, thus better decoding performance. Moreover, Soft decoding is outperforming the Hard one. Additionally, CISPM precoding is more secure than ZF, i.e. BER at Eve for CISPM is higher than the one of ZF. As for the use of the PLS random scheme, in Fig.~\ref{BER_vs_tSINR_13_3Ant}(b), we notice that, similarly, the higher the values of $\eta/\gamma_k$, the lower the BER. We also observe that soft decoding is the most performing with the difference being that using PLS random scheme offers much higher security compared to ZF and CISPM scheme, with high BER values at Eve even when using $11$ antennas at Eve. Concerning the PLS Eve-min-power scheme in Fig.~\ref{BER_vs_tSINR_13_3Ant}(c), it depicts the same behavior as PLS random. However, the latter scheme is more secure because of its incurred randomness in the precoding design, while PLS Eve-min-power scheme is designed with Eve's channel in the objective function that lowers the received power at Eve.

\begin{figure}[!htb]
\minipage{0.51\textwidth}
  \includegraphics[width=\linewidth]{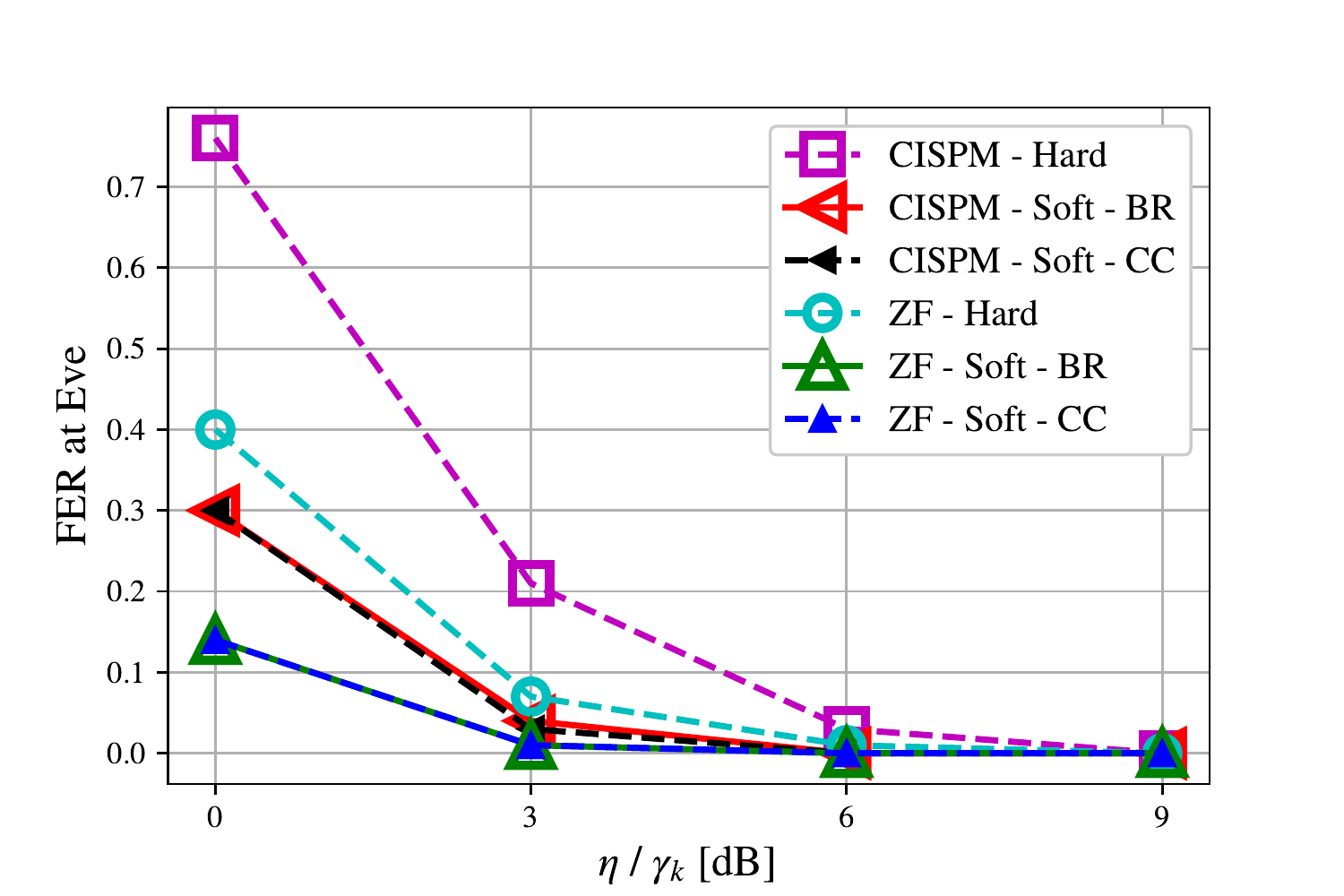}
  \caption*{(a) ZF and CISPM scheme}\label{fig:awesome_image1}
\endminipage\hfill
\minipage{0.51\textwidth}
  \includegraphics[width=\linewidth]{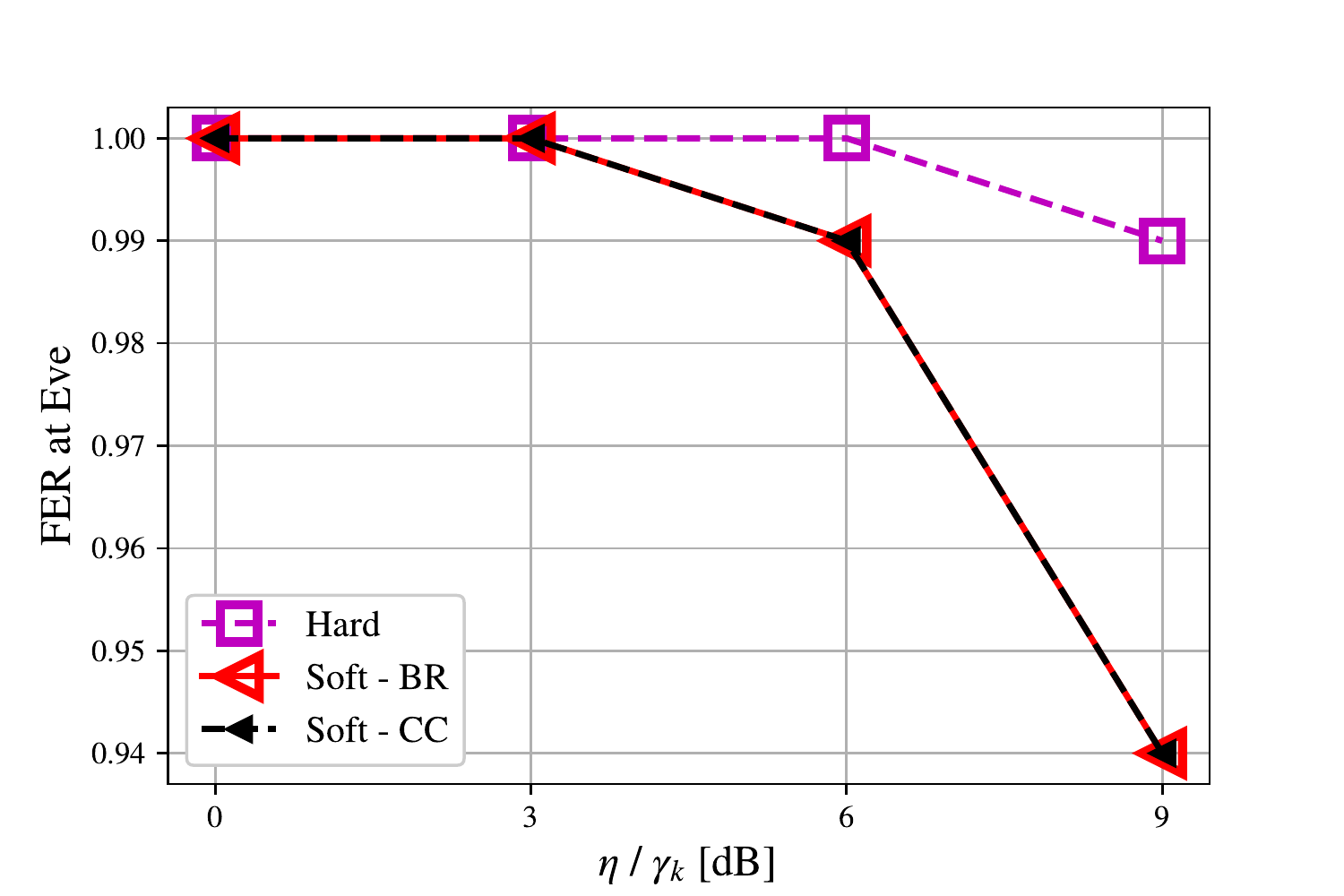}
  \caption*{(b) PLS random scheme}\label{fig:awesome_image2}
\endminipage\hfill
\minipage{0.51\textwidth}%
  \includegraphics[width=\linewidth]{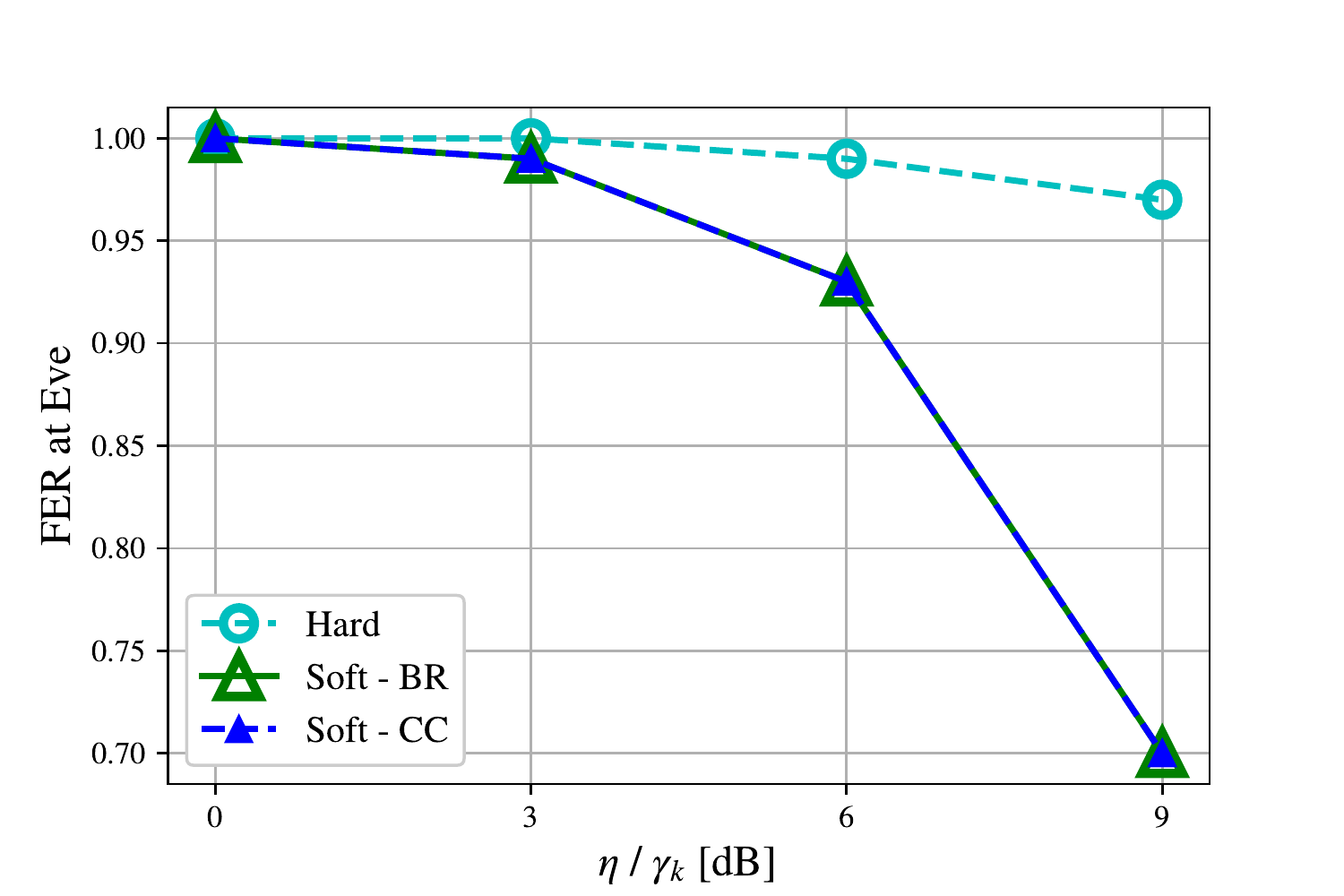}
  \caption*{(c) PLS Eve-min-power scheme}\label{fig:awesome_image3}
\endminipage
\caption{QPSK-coded FER at Eve vs. $\eta/\gamma_k$ $\mathrm{[dB]}$, with $r=\sfrac{1}{4}$ , $N_{\rm t} = 15$, $K = 6$, and $M = 11$.}
\label{FER_vs_tSINR_13_3Ant}
\end{figure}

Fig.~\ref{FER_vs_tSINR_13_3Ant} depicts the coded FER at Eve as a function of $\eta/\gamma_k$ $\mathrm{[dB]}$. The parameters used in the simulation are: $r=\sfrac{1}{4}$, $N_{\rm t} = 15$, $K = 6$, and $M = 11$. In Fig.~\ref{FER_vs_tSINR_13_3Ant}(a), for ZF and CISPM precoding, we notice that the higher the values of $\eta/\gamma_k$, the lower the FER, which is due to the increase of the transmit power. Particularly, soft decoding outperforms hard decoding, with FER values decently low, which validates the eavesdropping attack for ZF and CISPM precoding schemes, with CISPM being more secure. In Fig.~\ref{FER_vs_tSINR_13_3Ant}(b) however, when we use the PLS random scheme, we notice that the higher the values of $\eta/\gamma_k$, the lower the FER. Particularly, the high FER values validate the security of the PLS random scheme. Lastly, when using the PLS Eve-min-power scheme, in Fig.~\ref{FER_vs_tSINR_13_3Ant}(c), we observe the same behavior as in the case of the PLS random scheme, with FER values at Eve lower than ones for the PLS random approach. This validates the high security exhibited by the PLS random scheme, which outperforms the PLS Eve-min-power scheme's security performance. Yet, the FER values for the PLS Eve-min-power are still very high compared to the non-secure schemes, i.e., ZF and CISPM schemes.

 Next, we investigate the FER at user $k$ as a function of $\eta/\gamma_k$ $\mathrm{[dB]}$ by comparing ZF and CISPM schemes with the PLS schemes. The parameters used in the simulation setup are: $r=\sfrac{1}{3}$, $N_{\rm t} = 15$, $K = 6$, and $M = 11$. We found out that the values of the FER at user $k$ are all zeros for all of the schemes even when $M = 11$, which was the reason to omit the plot in the manuscript. With such a low code rate, we do not obtain a single error for the entire range of $\eta/\gamma_k \in \{0, 1, 2, 3\}$. This happens in both schemes since we align the transmitted signal to the intended users' channels, which in turn will receive their intended symbols in the corresponding detection regions. Thus, the proposed PLS schemes provide much higher security without impacting the intended user's performance. 

\begin{figure}[t]
\centering
  \includegraphics[width=3.5in]{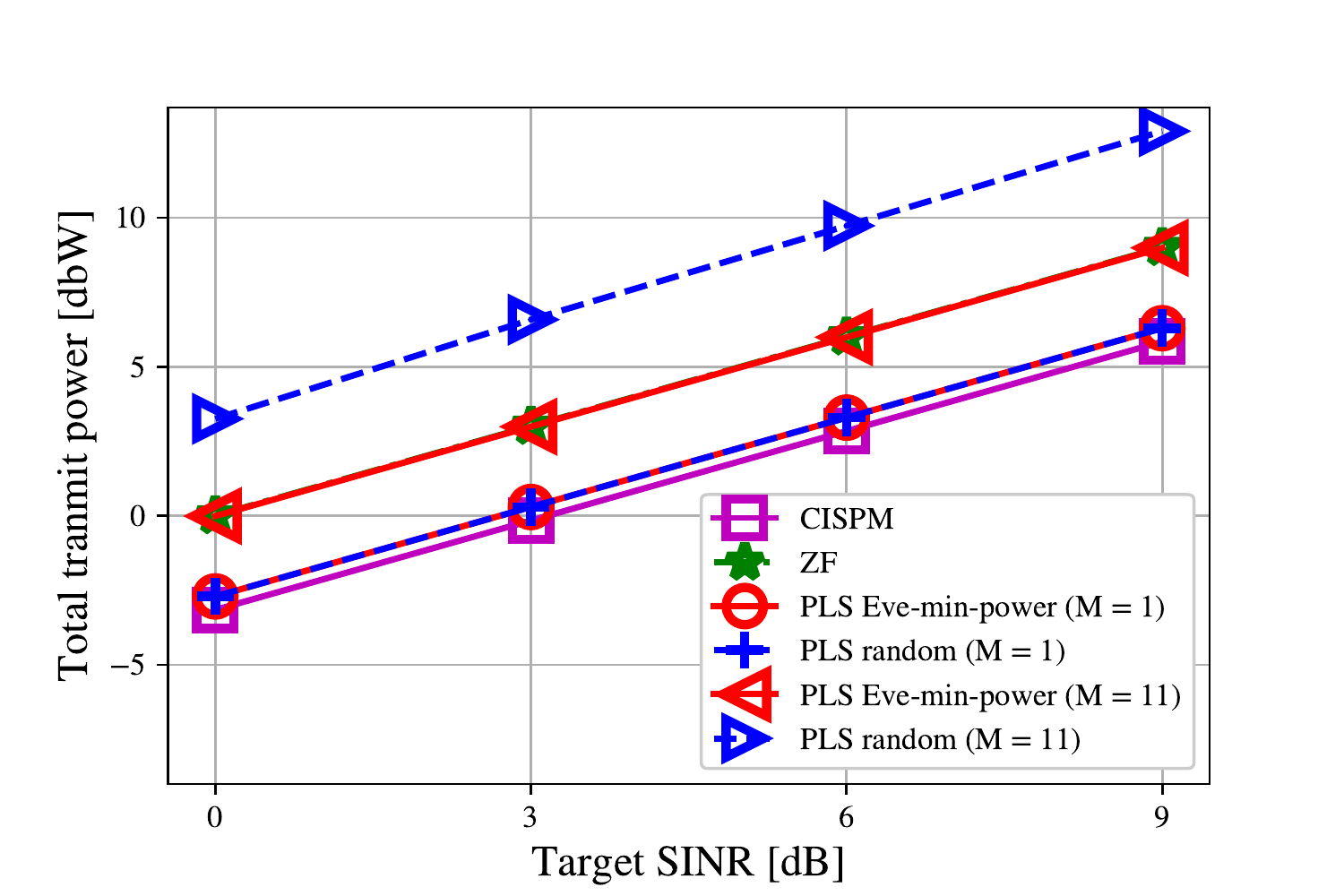}\\ 
  \caption{QPSK-Total tansmit power [dBW] vs. target SINR [dB], with $N_{\rm t} = 15$, $K = 6$, and $M \in \{ 1, 11 \}$.}
  \label{avePower_vs_tSINR}
\end{figure} 

Fig.~\ref{avePower_vs_tSINR} shows the total transmit power $P_{\rm tot}$, in dBW, as a function of $\eta/\gamma_k$, which we set to the same value for all users for simplicity. We compare ZF and CISPM schemes with the PLS ones. The parameters used in the simulation are: $N_{\rm t} = 15$, $K = 6$, and $M = \{1,11\}$. As explained above, the higher the $\eta/\gamma_k$ values, the higher is the transmit power, for all the schemes. For the CISPM scheme, we observe that $P_{\rm tot}$ is lower than the target SINR at the intended users, which is due to the constructive interference turning into power gains at the receivers, hence less transmit power is required to attain the desired SINR value. However for ZF precoding, as predicted, $P_{\rm tot}$ is the same as the mean power $\eta$, as it has been set. As for the PLS schemes, $P_{\rm tot}$ depends on the number of antennas at Eve $M$, higher $M$ leads to higher $P_{\rm tot}$. This increase is due to the fact that more antennas at Eve imply more constraints in the case of PLS random that result in the observed big increase in $P_{\rm tot}$ for $M = 11$. To elaborate more, this behavior is due to the fact that, the more we constrain our signal design problem, the more power is required to solve it. However, in the case of PLS Eve-min-power, the higher $M$, the higher the power consumption, i.e., the more antennas at Eve, the Eve-related part of the objective function tends to have higher values due to the higher degrees of freedom at the Eve's side, thus higher power consumption; even for $M = 11$, PLS Eve-min-power does not consume as much power as PLS random, its consumption is in fact equivalent to ZF scheme in $P_{\rm tot}$. However, for $M = 1$, the two proposed PLS schemes consume the same power. 

We conclude this section by summarizing the insights from the numerical results. 
\begin{enumerate}
    \item Soft decoding schemes always outperform hard decoding, i.e., soft values carry extra information that is used by the decoder to better estimate the original data. 
    \item Soft – CC and Soft – BR performance is the same because of the lack of label-correlation, due to the random nature of data to be transmitted. 
    \item CISPM precoding is more secure than ZF because the precoding pattern changes at each symbol-period while ZF precoding is fixed throughout the whole coherence time. 
    \item Proposed PLS schemes are much more secure than CISPM and ZF, with PLS random being the most secure because of its induced randomness in the signal design that makes it harder for Eve to learn the precoding pattern. 
    \item Logistic regression is the most performing classifier amongst the tested state-of-the-art classifiers.
    \item The system parameters that directly affect the BER/FER at Eve are: the number of antennas at Eve, the total transmit power, and the coding rate. 
    \item PLS schemes offer significant security gains compared to ZF and CISPM precoding schemes at the expense of additional power consumption at the transmitter. 
    \item More importantly, these security gains are achieved without affecting the performance at the intended users. 
    \end{enumerate}

\section{Conclusions}\label{CC}
In this paper, we proposed ML-based decoding schemes for a multi-antenna Eve in the context of a FEC-enabled MU-MISO systems. The proposed eavesdropping attacks use precoded pilot symbols as training data and enable an Eve to soft/hard decode a message with high accuracy. As a countermeasure to these attacks, we proposed two novel security-enhanced SLP precoders that seek to obstruct the learning process at Eve. Numerical results validated both the attacks as well as the countermeasures, where the soft decoding scheme always outperforms the hard decoding one. In addition, our proposed PLS schemes outperform ZF and CISPM precoding in security at the expense of additional power consumption at the transmitter, with PLS random scheme offering the highest security. Thus, the proposed PLS schemes provide different trade-offs between security and power consumption, which would give the BS the option to select the most suited scheme depending on the required criteria. Notably, despite all the security gains offered by our proposed PLS schemes, their use does not affect the performance at the intended user. Future research topics would be to extend this work to the case of imperfect CSI and also where the channel to Eve is unknown to the BS. 
 
\bibliographystyle{IEEEtran}

\addcontentsline{toc}{section}{\refname}
\bibliography{IEEEabrv,mybib}

\end{document}